\documentclass[twocolumn]{aastex63}
\usepackage{longtable}
\usepackage{CJKutf8, here}

\newcommand{\kanji}[1]{\begin{CJK}{UTF8}{ipxm}(#1)\end{CJK}}

\received{---}
\revised{---}
\accepted{---}
\submitjournal{ApJ}

\shorttitle{Effects of Nuclear EOS on the Type-I X-ray Bursts}
\shortauthors{Dohi et al.}

\begin{document}


\title{Effects of the Nuclear Equation of State on Type I X-Ray Bursts: Interpretation of the X-Ray Bursts from GS~1826-24}

\correspondingauthor{Akira Dohi}
\email{dohi@phys.kyushu-u.ac.jp}

\author[0000-0001-8726-5762]{A.~Dohi \kanji{土肥明}}
\affiliation{Department of Physics, Kyushu University, Fukuoka 819-0395, Japan}
\affiliation{Interdisciplinary Theoretical and Mathematical Sciences Program (iTHEMS), RIKEN, Wako, Saitama 351-0198, Japan}

\author[0000-0002-0842-7856]{N.~Nishimura \kanji{西村信哉}}
\affiliation{Astrophysical Big Bang Laboratory, Cluster for Pioneering Research, RIKEN, Wako, Saitama 351-0198, Japan}
\affiliation{RIKEN Nishina Center for Accelerator-Based Science, Wako, Saitama 351-0198, Japan}
\affiliation{Division of Science, National Astronomical Observatory of Japan, Mitaka 181-8588, Japan}

\author{M.~Hashimoto \kanji{橋本正章}}
\author{Y.~Matsuo \kanji{松尾康秀}}
\affiliation{Department of Physics, Kyushu University, Fukuoka 819-0395, Japan}

\author[0000-0003-0943-3809]{T.~Noda \kanji{野田常雄}}
\affiliation{Department of Education and Creation Engineering, Kurume Institute of Technology, Kurume, Fukuoka 830-0052, Japan}

\author[0000-0002-7025-284X]{S.~Nagataki \kanji{長瀧重博}}
\affiliation{Astrophysical Big Bang Laboratory, Cluster for Pioneering Research, RIKEN, Wako, Saitama 351-0198, Japan}
\affiliation{Interdisciplinary Theoretical and Mathematical Sciences Program (iTHEMS), RIKEN, Wako, Saitama 351-0198, Japan}

\begin{abstract}
Type I X-ray bursts are thermonuclear explosions on the neutron star (NS) surface caused by mass accretion from a companion star. Observations of X-ray bursts provide valuable information on X-ray binary systems, e.g., binary parameters, the chemical composition of accreted matter, and the nuclear equation of state (EOS) of NSs. There have been several theoretical studies to constrain the physics of X-ray bursters. However, they have mainly focused on the burning layers above the solid crust of the NS, which brings up issues of the treatment of NS gravitation and internal energy. In this study, focusing on the microphysics inside NSs, we calculate a series of X-ray bursts using a general-relativistic stellar-evolution code with several NS EOSs. We compare the X-ray-burst models with the burst parameters of a clocked burster associated with GS~1826-24. We find a monotonic correlation between the NS radius and the light-curve profile. A larger radius shows a higher recurrence time and a large peak luminosity. In contrast, the dependence of light curves on the NS mass becomes more complicated, where the neutrino cooling suppresses the efficiency of nuclear ignition. We also constrain the EOS and mass of GS~1826-24, i.e., stiffer EOSs, corresponding to larger NS radii, are not preferred due to too-high peak luminosity. The EOS and the cooling and heating of NSs are important to discuss the theoretical and observational properties of X-ray bursts.
\end{abstract}


\section{Introduction} \label{sec:intro}

Type I X-ray bursts (simply X-ray burst\footnote{In the present paper, we investigate ``standard'' type I X-ray bursts and do not mention superbursts (i.e., a subclass of the type I X-ray bursts) and type II X-ray bursts.}, hereafter) are transients observed in X-ray binaries, the luminosity of which increases by one order of magnitude over the duration of hours to days \citep[e.g.,][]{2006NuPhA.777..601S, 2013PrPNP..69..225P}. The energy source of the X-ray bursts is nuclear burning on the surface of neutron stars (NSs), where accreted hydrogen and helium from a companion star are converted into heavy nuclei initiated by the triple-$\alpha$ reaction and the following hot-CNO cycle. In such proton-rich explosive environments, the rapid-proton-capture process (rp-process) can occur \citep{1998PhR...294..167S, 2001PhRvL..86.3471S}. Since their first discovery in the mid-1970s \citep{1976ApJ...206L.135B, 1976ApJ...205L.127G}, thousands of burst events from over 100 sources (bursters) have been identified by several X-ray satellites \citep[see, e.g.,][]{2020ApJS..249...32G}. These observations can be a probe for investigating the physical environments of X-ray binaries and properties of NSs \citep[for a review, see][]{2021ASSL..461..209G}.

As there has been no clear observational evidence of their compositions in nuclear ashes, light curves are almost the only observational probe for X-ray bursts. The physical mechanism of X-ray bursts has been investigated in many works \citep[e.g.,][]{1987ApJ...319..902F, 2004ApJS..151...75W, 2012ApJ...752..150K} using  one-dimensional burst simulations assuming spherical symmetry. They found some input parameters are of importance to describe the burst models: the mass accretion rate ($\dot{M}$), the initial metallicity ($Z_{\rm CNO}$) in the accreted matter, the hydrogen to helium ratio ($X/Y$), the reaction rates of the light nuclei, and the rp-process path. On the other hand, burst light-curve properties are characterized by a set of {\it output} parameters: the recurrence time ($\Delta t$) between burst events, the peak luminosity ($L_{\rm peak}$), the burst duration from the peak ($\tau$), and the burst strength ($\alpha$), which is the ratio of the burst flux to the persistent flux.

The observations of light curves, in general, do not strongly constrain X-ray-burst model parameters such as $\dot{M}$ and $Z_{\rm CNO}$. However, specific cases, called clocked bursters, which show a similar burst profile in each epoch \citep[for a review, see][]{2017PASA...34...19G}, provide strict constraints on theoretical models. GS1826-24, first observed in 1989 by the Ginga satellite \citep{1989ESASP.296....3T}, is a representative clocked burster \citep[e.g.,][]{1999ApJ...514L..27U}. By comparing the light curve of clocked bursters with numerical models, burst parameters can be derived. These are several studies that estimated the initial metallicity and the accretion rate of GS1826-24 \citep{2007ApJ...671L.141H, 2018ApJ...860..147M}. Recently, \cite{2020MNRAS.494.4576J}, based on 3840 burst models, constrained some model parameters such as $Z_{\rm CNO}=0.01^{+0.005}_{-0.004}$, and $\dot{M}=0.084$--$0.170~\dot{M}_{\rm Edd}$, where $\dot{M}_{\rm Edd}=1.75\times10^{-8}~M_{\odot}~{\rm yr}^{-1}$ is the typical Eddington accretion rate on the NS. The impacts of nuclear reaction rate uncertainties on X-ray-burst models have also been discussed \citep[][]{2019ApJ...872...84M}.

However, their burst models using {\tt KEPLER}~\citep[e.g.,][]{2020MNRAS.494.4576J} and {\tt MESA}~\citep{2018ApJ...860..147M,2019ApJ...872...84M} cover above the NS solid crust layer, where the inner boundary corresponds to the crust surface. This brings up the following two issues (but see the burst models of, e.g., \cite{2016ApJ...831...13D}, using dStar \cite{2015ascl.soft05034B} as an exception). One is the treatment of NS gravitation, which is based on nonrelativistic hydrodynamic formulation. Although sophisticated correction of general relativity has been well considered with the use of the gravitational-redshifted factor~\citep{2011ApJ...743..189K}, burst models based on a relativistic formulation are primarily desired for more exact calculation of light curves. The other one is to give the boundary condition on the crust surface. In other words, they necessarily treat the boundary luminosity or temperature, which is artificially given without fully considering the microphysics inside NSs, e.g., the crustal heating, the neutrino emission, and the uncertainty of the nuclear equation of state (EOS). The heating and neutrino cooling of the NS, which affect the burst light curves via the change in the surface temperature, are essential for X-ray burst modelling \citep[see, e.g.,][]{1984ApJ...278..813F,1987ApJ...319..902F,2006ApJ...646..429C}. To include the energy exchange between the interior NS and the accreting layer, previous studies adopted a heating factor $Q_b$ on the crust in the energy equation, which was treated as an adjustment parameter. Although $Q_b<0.5~{\rm MeV~u^{-1}}$ is implied \citep[][]{2018ApJ...860..147M,2020MNRAS.494.4576J} by comparison to the light-curve observation, this value essentially should be determined by solving the thermal evolution of the NS.

Besides the NS heating and cooling, the EOS plays an important role in X-ray-burst modeling, because the surface gravity affects the amount of fuel \citep[][]{1981ApJ...247..267F}. Recent observations by NICER provide additional constraints on the mass and radius of NSs \citep{2019ApJ...887L..22R, 2019ApJ...887L..24M}. The observation of gravitational waves from NS mergers, e.g., GW170817 \citep{2018PhRvL.121p1101A}, GW190425 \citep{2020ApJ...892L...3A}, and GW190814 \citep{2020ApJ...896L..44A}, give another restriction on the EOS. Observations of X-ray bursts, in particular, clocked bursters (e.g., GS1826-24), can be an alternative probe to constrain the EOS. Similar to the treatment of $Q_b$, the general-relativistic effect should be considered by solving X-ray-burst calculations over whole NS regions as well.

In this study, we investigate the effects of the NS EOS on the X-ray-burst models. We employ a general-relativistic stellar-evolution code with the latest microphysics \cite[i.e., the EOS and cooling and heating effects;][]{2020PTEP.2020c3E02D}, considering the NS from the central core to the regions of the atmosphere. Adopting several recent EOSs, we examine the mass of NS GS~1826-24 by comparing between theoretical models and the observational X-ray-burst parameters. We discuss the impacts of the uncertainty of the EOS on the mass determination of GS~1826-24, which can be a new observational constraint on the NS EOS. We also emphasize the importance of cooling and heating sources inside the NS for the X-ray-burst modeling.

This paper is organized as follows. In Section~\ref{sec:method}, we describe the basic formulae, numerical method, and input parameters. In Section~\ref{sec:init}, we present the initial models of burst simulations. In Section~\ref{sec:results}, the results of the light curves and characteristic parameters are shown. We discuss the physical uncertainties of the NS interiors and their effects on the burst parameters. We constrain the EOS and mass using GS1826-24. Section~\ref{sec:con} is devoted to the conclusion.

\section{Methods}\label{sec:method}

To solve the evolution of X-ray-burst models including the central NS, we use a one-dimensional general-relativistic stellar-evolution code \citep{1984ApJ...278..813F} with the realistic nuclear EOSs (see \cite{2020PTEP.2020c3E02D} for details). We describe the basic equations and microphysics considered in the following X-ray-burst calculations.

\begin{figure}[tbp]
    \centering
    \includegraphics[width=\linewidth]{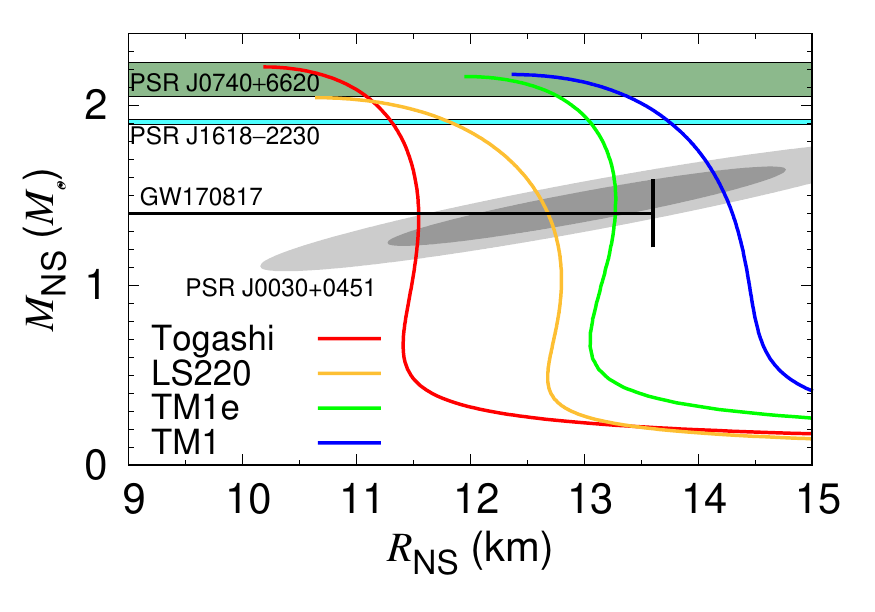}
    \caption{The mass--radius relations of NSs with several EOSs.  We highlight the mass range of three observed NSs: $M_{\rm NS}=1.908\pm0.016~M_{\odot}$ for PSR J1618-2230~\citep{2010Natur.467.1081D,2018ApJS..235...37A} and $M_{\rm NS}=2.14^{+0.10}_{-0.09}~M_{\odot}$ for PSR J0740+6620~\citep{2020NatAs...4...72C}. The range of the NS radius constrained by GW170817 \citep{2018PhRvL.121p1101A} is shown, i.e., $R_{1.4}<13.6~{\rm km}$ \citep{2018PhRvL.120q2703A, 2018PhRvL.120z1103M}, where $R_{1.4}$ is the radius of $1.4~M_{\odot}$ stars. Another observational constraint by NICER \citep{2019ApJ...887L..24M} from PSR J0030+0451 \citep[][]{2020Univ....6...81B} is also highlighted.}
    \label{fig:mr}
\end{figure}

\subsection{The basic equations}

We assume that hydrostatic equilibrium is achieved in the X-ray-burst system, covering from the central NS core to the mass accreting layer on the surface (for the basic equations, see \cite{2020PTEP.2020c3E02D}). In the accreted layer, we utilize the Eulerian coordinate of the mass fraction $q \equiv M(r)/M(t)$, where $M(r)$ is the rest mass enclosed inside the radius $r$ and $M(t)$ is the total mass at each time. For stellar-evolution calculations, the variable $q$ is more useful than $M(r)$, because the computational time becomes much shorter \citep[][]{1981PThPS..70..115S}. The gravitational energy release rate, depending on on time, is divided into two terms, i.e., the nonhomologous ($\varepsilon_{\rm g}^{nh}$) and the homologous ($\varepsilon_{\rm g}^{h}$) components:
\begin{eqnarray}
\varepsilon_{\rm g}^{nh} &=& e^{-\phi/c^2}\left(\left.T\frac{\partial s}{\partial t}\right|_q + \mu_i\left.\frac{\partial N_i}{\partial t}\right|_q\right), \label{eq:g1}\\
\varepsilon_{\rm g}^{h} &=& e^{-\phi/c^2}\dot{M}\left(\left.T\frac{\partial s}{\partial \ln(q)}\right|_t + \mu_i\left.\frac{\partial N_i}{\partial \ln(q)}\right|_i\right), \label{eq:g2}
\end{eqnarray}
where $\phi$ is the gravitational potential, $s$ is the specific entropy, and $\mu_i$ and $N_i$ are the chemical potential and the number per unit mass of the $i$th elements, respectively. The latter term $\varepsilon_{\rm g}^{h}$ indicates the compressional heating due to the accretion, which significantly contributes as a heat source in some cases \citep[e.g.,][]{2017JPSJ...86l3901L}. 

In addition, the EOS and the opacity are necessary to close the hydrostatic equation system (we describe details of the EOS in the next section). The opacity includes the components of radiation and conductive opacity \citep{1999ApJ...524.1014S, 2001A&A...374..151B, 2015SSRv..191..239P}\footnote{To calculate the electron conductivity, we use a numerical code, available in \url{http://www.ioffe.ru/astro/conduct/}.}. For the NS surface of the accreted layer, we assume the radiative zero boundary condition \citep{1984ApJ...278..813F}. As the outer-mesh points need to be sufficiently close to the photospheric area, we set the outer-mesh mass $M_{\rm outer}=10^{-19}M_{\rm NS}$.

Using the above conditions, we solve these hydrostatic equations from the center to the surface using the Henyey-type numerical scheme of the implicit-midpoint method with an adaptive grid in $q$ and $t$. With the nuclear composition varying due to nuclear burning, we calculate the density and temperature in the hydrostatic stellar structure each time step. To follow the nuclear burning process, we solve the nuclear reaction network (see Section~\ref{sec:network}) consistent with teh stellar structure. By continuing this procedure for each time, we obtain time evolution of the bolometric luminosity of accreting neutron stars.

\begin{figure}[h]
    \centering
    \includegraphics[width=\linewidth]{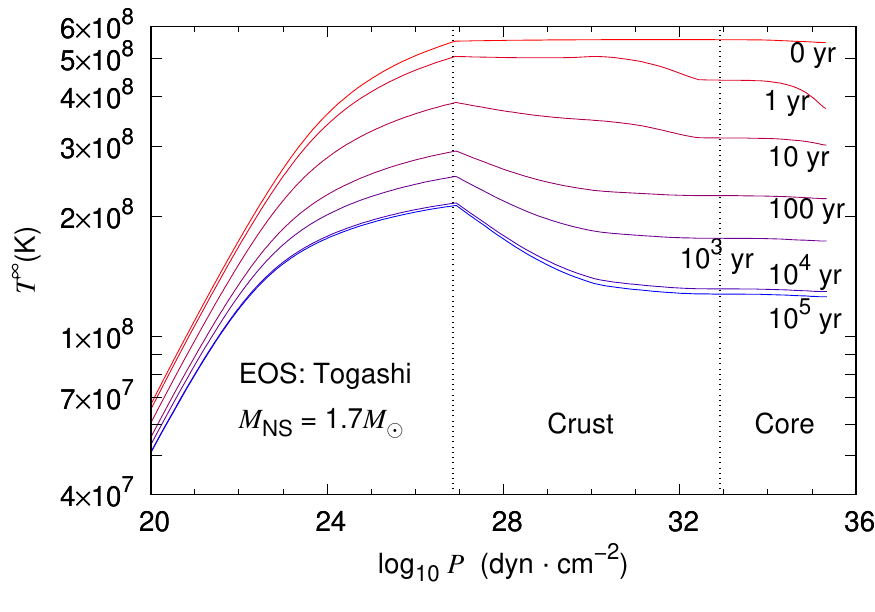}
    \caption{The redshifted temperature for a $1.7~M_{\odot}$ NS with the Togashi EOS with the mass accretion of $\dot{M}_{-9}=2.5$. The time snapshots from $0$~yr to $10^5$~yr are plotted, where the initial state satisfies the isothermal condition. The effect of nuclear heating is ignored. Dotted lines indicate the pressure $P\simeq7\times10^{26}~{\rm dyn~cm^{-2}}$ on the crust surface and $P \simeq 8 \times 10^{32}~{\rm dyn~cm^{-2}}$, which corresponds to the nuclear saturation density ($ = 2.66 \times 10^{14}~{\rm g~cm^{-3}}$).}
    \label{fig:quies}
\end{figure}

\subsection{The EOS and heating and cooling}

In the present study, we adopt four sets of finite-temperature EOSs: Togashi \citep{2017NuPhA.961...78T}, LS220 \citep{1991NuPhA.535..331L}, TM1 \citep{1998NuPhA.637..435S, 1998PThPh.100.1013S, 2011ApJS..197...20S}, and TM1e \citep{2019ApJ...887..110S, 2020ApJ...891..148S}.  While many astrophysical simulations have widely used the LS220 and TM1 EOSs, the Togashi and TM1e EOSs are recently constructed: the Togashi EOS is based on the bare nuclear force for two-body interaction and phenomenological three-body interaction. The TM1e EOS is based on the extended relativistic mean-field theory updated from the TM1 EOS, and the $\omega$--$\rho$ coupling term is added in the nucleonic Lagrangian density, which significantly affects the softness. For low-density regions around the crust surface, we connect the nuclear EOSs to the ones in \cite{1971ApJ...170..299B} and \cite{1982ApJ...255..624R}.

In Fig.~\ref{fig:mr}, we show the mass-radius relation for each EOS. All EOSs satisfy observational constraints to the maximum mass of NSs, whose value is currently $M_{\rm NS} = 2.14^{+0.10}_{-0.09} M_{\odot}$ \citep[][]{2020NatAs...4...72C}. Moreover, we can see that the adopted EOSs in this work have different radii, which basically depend on the difference in the symmetry energy. There are several constraints on the radius, such as GW170817 and PSR J0030+0451 from NICER. Their estimated values are in the wide range of around $12$--$14$ km, with which Togashi, LS220, and TM1e are in good agreement.

Such a uncertainty of radius deduced from EOS affects not only the amount of fuel in outburst but also the strength of the heating and cooling which occur in the NS crust and core. As the heating process inside NS, we introduce the crustal heating, which is the non-equilibrium nuclear reactions caused by the compression of the matter falling on the NS crust surface. Specifically, we adopt classical model of \cite{1990A&A...227..431H}, and the total produced energy is around $1.4~{\rm MeV}$ per one accreted nucleon. There are some developed versions of crustal heating models~\citep[e.g.][]{2018A&A...620A.105F}, but according to \cite{2008A&A...480..459H}, the heating rate of \cite{1990A&A...227..431H} is shown to be quite a reasonable estimate. For the cooling process inside NS, the neutrino emission occurring inside NS through weak interactions should be considered. There are many kinds of neutrino emission processes, including nucleon superfluid effect on thermal evolution of NSs~\citep[for review, see][]{2001PhR...354....1Y,2013arXiv1302.6626P}, but for simply, we adopt the slow neutrino cooling processes including the modified Urca process and bremsstrahlung. For LS220 and TM1 EOSs, since their values of nucleonic symmetry energy are high enough to cause the Direct Urca process, such a fast cooling process occurs with relatively low-mass stars~\citep[e.g.,][]{2017IJMPE..2650015L,2019PTEP.2019k3E01D}. In this work, however, to focus on the difference of the radius as the EOS dependence, we do not incorporate it.

\subsection{The approximate nuclear reaction network}
\label{sec:network}

We use an approximate network (APRX), which can reproduce the nuclear energy generation rates and nucleosynthesis products in X-ray-burst conditions. The construction of the approximate reaction networks is based on the calculations of the nucleosynthesis by full reaction networks \citep{2003ApJ...585..418F,2004ApJ...603..242K}, covering all relevant reactions in the proton-rich area. To reduce the numerical cost effectively, we adopt a reduced network with 88 nuclei \cite[for details, see Table 1 in][and references therein]{2020PTEP.2020c3E02D}. We take the nuclear reaction rates mostly from the JINA Reaclib (ver 2.0), \citep[][]{2010ApJS..189..240C}\footnote{https://jinaweb.org/reaclib/db/}.

The APRX network is 150 times faster than the large reaction networks with 897 nuclei~\citep{2017KUPhD1806813}. Moreover, the APRX is shown to reproduce the energy generation rate and hydrogen mass fraction with a large reaction network with 897 nuclei according to \cite{2020PTEP.2020c3E02D}. Thus, the APRX is useful in terms of the saving in numerical computation as well as the sufficient reproducibility of the full nuclear reaction network.

\begin{figure}[t]
    \centering
    \begin{minipage}{\linewidth}
    \includegraphics[width=\linewidth]{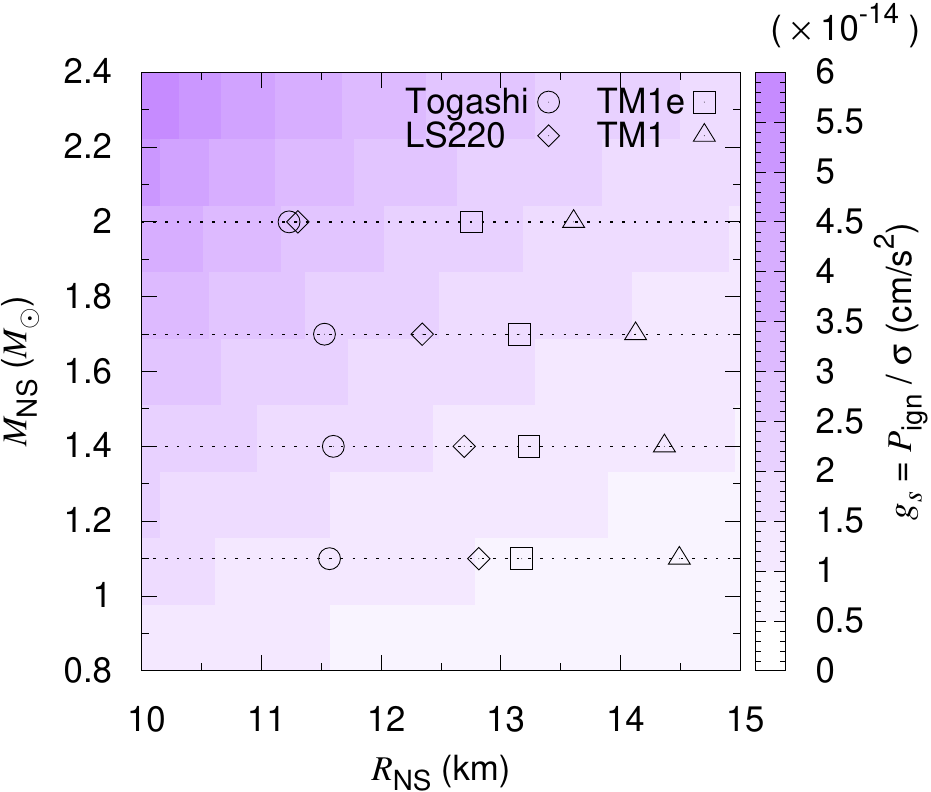}
    \end{minipage}
    \caption{The surface gravity acceleration $g_{\rm s}$ on the mass-radius plane, based on the (one-zone) shell-flash model \citep{1981ApJ...247..267F}. The symbols indicate the values of our multizone X-ray-burst calculations.}
    \label{fig:FHM}
\end{figure}

\begin{figure}[h]
    \centering
     \includegraphics{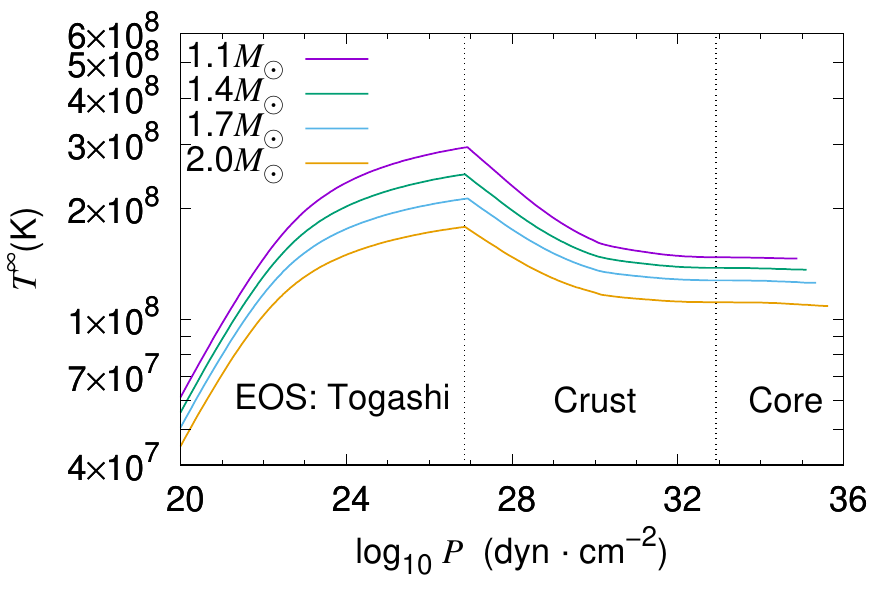}
     \caption{The redshifted temperature structure in the steady state with $M_{\rm NS}=1.1,1.4,1.7,~{\rm and}, 2.0~M_{\odot}$, assuming the Togashi EOS and $\dot{M}_{-9}=2.5$. Same as Fig.~\ref{fig:quies} for dotted lines.}
     \label{fig:quies_mass}
\end{figure}

\section{The initial models for burst calculations}
\label{sec:init}

\subsection{Preburst evolution without nuclear burning}

To develop the initial conditions for the X-ray-burst simulations, we first calculate the thermal evolution of accreting NSs without nuclear burning from the isothermal state \citep{2020PTEP.2020c3E02D}. In Fig.~\ref{fig:quies}, we show the time evolution of the redshifted temperature without nuclear burning from $0$~yr at the beginning of the calculation to $10^5$~yr at the end of the preevolution. We adopt the mass accretion rate of $\dot{M}_{-9}=2.5$ for $1.7~M_{\odot}$ stars with the Togashi EOS, where $\dot{M}_{-9}$ is the accretion rate normalized by $10^{-9}~M_{\odot}~{\rm yr}^{-1}$. The initial state at $t=0~{\rm yr}$ is constructed to satisfy the isothermal condition.

We cover the entire region from the core to the accreted layers in the numerical calculation. Although some previous thermal evolution models of accreting NSs fix the temperature in the core~\citep[e.g.,][]{2019ApJ...882...91L}, we do not keep it fixed. Due to the neutrino emission from the crust and core of the NS, the temperature decreases with time. However, the temperature structure settles in the steady state at $t=10^4$--$10^5~{\rm yr}$, because the effect of crustal heating accompanying with the mass accretion in the crust regions becomes significant. Note that we find that the temperature structure bends around $P \sim 10^{27}~{\rm dyn~cm^{-2}}$, which appears to be due to the switching of EOSs around the crust surface. We use such a quiescent NS model as the initial model for X-ray-burst calculation.

\begin{figure}[t]
    \centering
    \includegraphics[width=\linewidth]{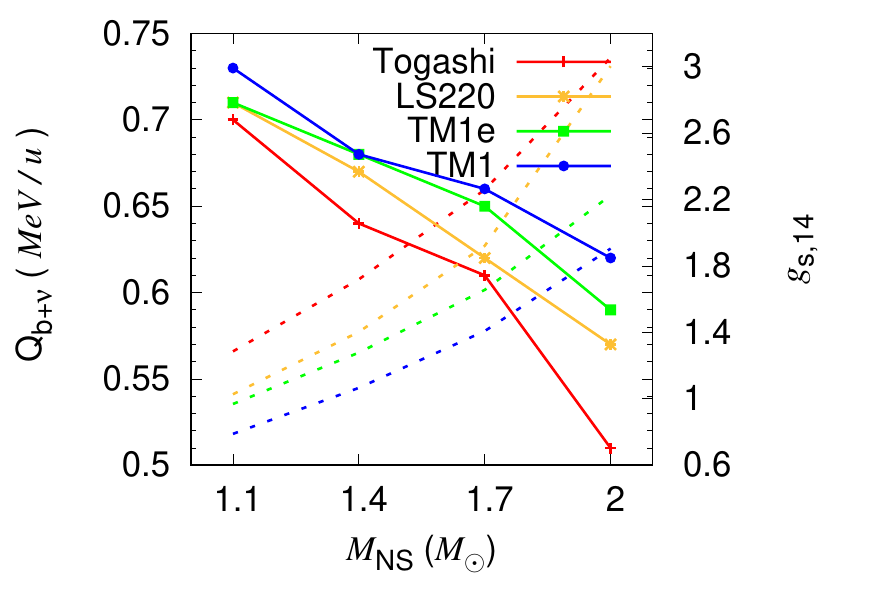}
    \caption{The $Q_{{\rm b} + \nu}$ (the solid lines for the left axis) and the surface gravitational acceleration $g_{{\rm s},14}$ (the dashed lines for the right axis) against $M_{\rm NS}$.}
    \label{fig:Q-gs}
\end{figure}

\begin{table}[t]
    \centering
     \caption{Calculated Values of $Q_{b+\nu}~[{\rm MeV~u^{-1}}]$}
    \begin{tabular}{ccccc}
    \hline\hline
    $M_{\rm NS}$     &  Togashi & LS220 & TM1e & TM1 \\
         \hline
    $1.1 M_{\odot}$     & 0.70 & 0.71 & 0.71 & 0.73 \\
    $1.4 M_{\odot}$     & 0.64 & 0.67 & 0.68 & 0.68 \\
    $1.7 M_{\odot}$     & 0.61 & 0.62 & 0.65 & 0.66 \\
    $2.0 M_{\odot}$     & 0.54 & 0.57 & 0.59 & 0.62 \\
    \hline
    \end{tabular}
    \label{tab:Q_beff}
\end{table}

\begin{figure*}[h]
    \centering
    \includegraphics[width=0.8\linewidth]{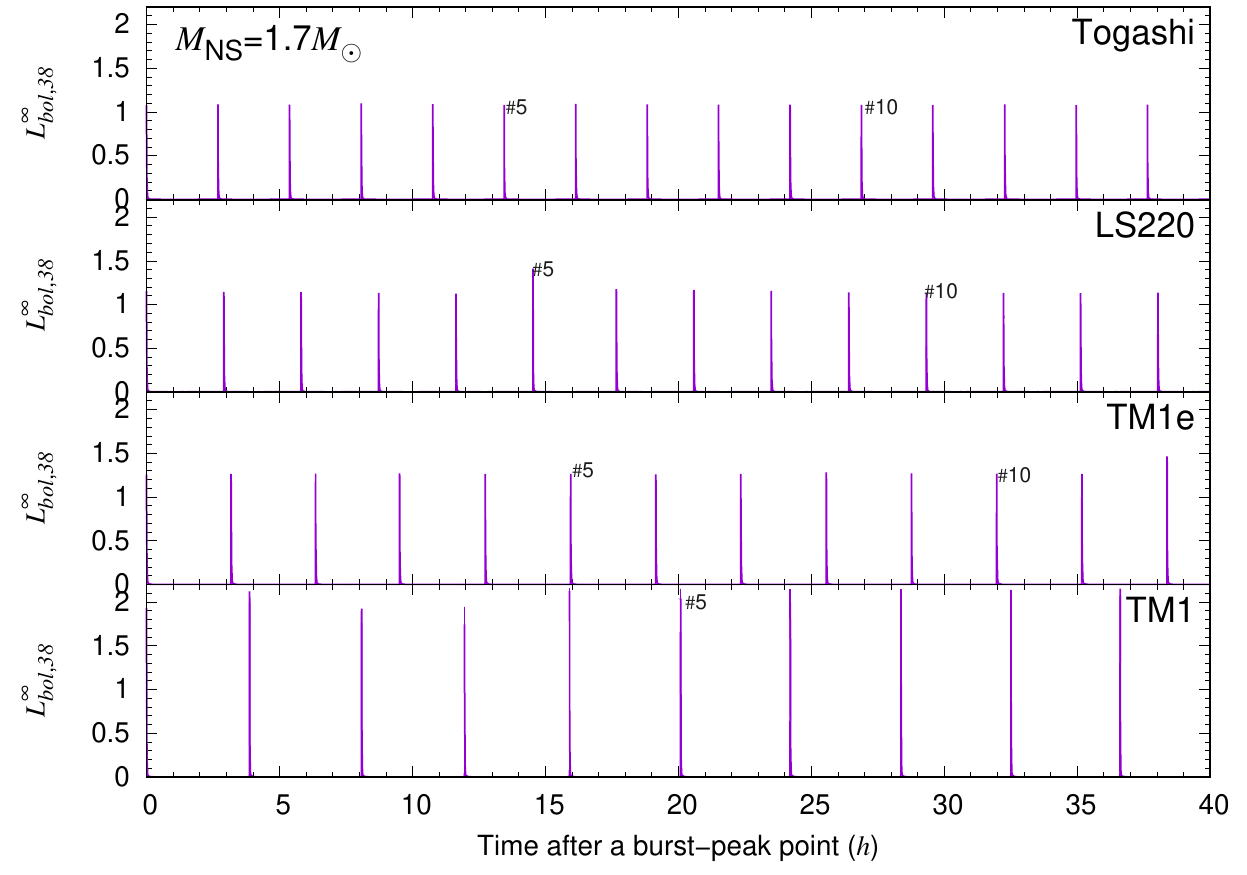}
    \caption{The luminosity of the burst sequence from $0$--$40~{\rm hr}$ for several NS EOSs. We adopt $M_{\rm NS}=1.7M_{\odot}$, $\dot{M}_{-9}=2.5$, and $Z_{\rm CNO} = 0.01$.}
    \label{fig:Lall_EOS}
\end{figure*}
\begin{figure*}[h]
    \centering
    \includegraphics[width=0.8\linewidth]{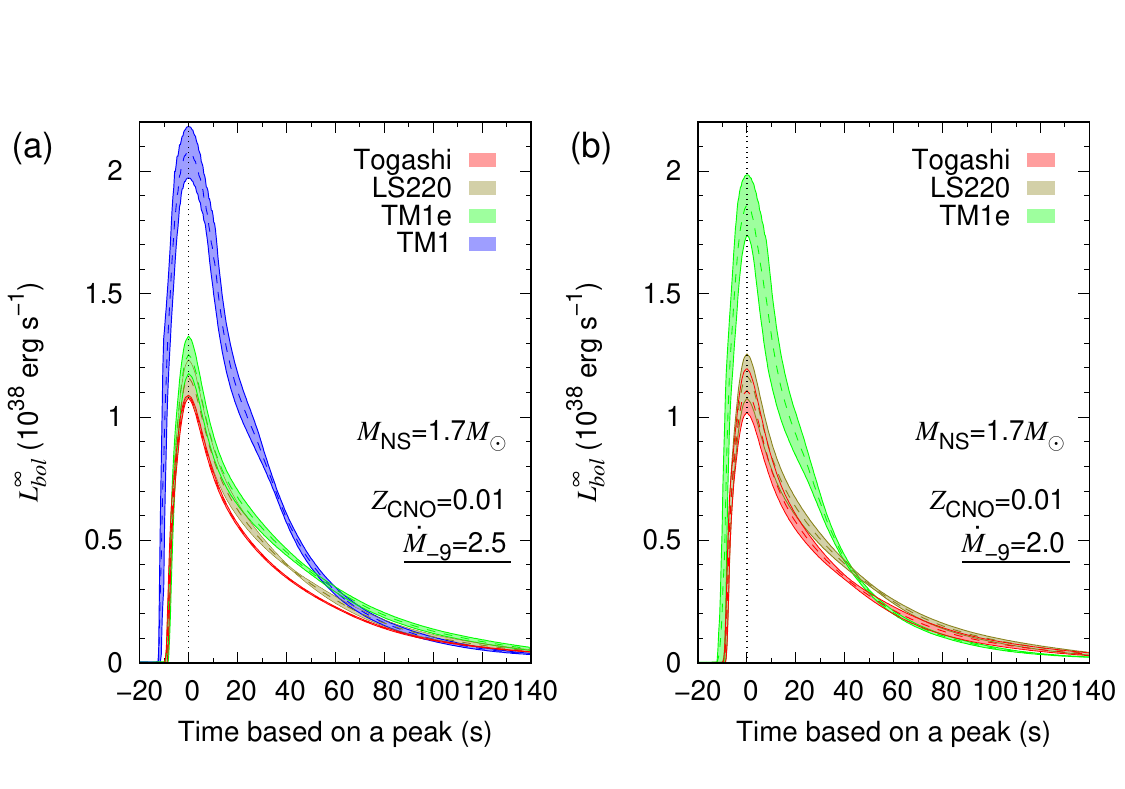}
    \caption{The light curves of the burst phase with $M_{\rm NS}=1.7 M_{\odot}$ for several EOSs. Assuming $Z_{\rm CNO}=0.01$, we adopt (a) $\dot{M}_{-9}=2.0$ and (b) $\dot{M}_{-9}=2.5$. We omitted the case of TM1 with $\dot{M}_{-9}=2.0$, which reaches the PRE.}
    \label{fig:Lb_EOS}
\end{figure*}

\begin{figure*}[h]
    \centering
    \includegraphics[width=0.8\linewidth]{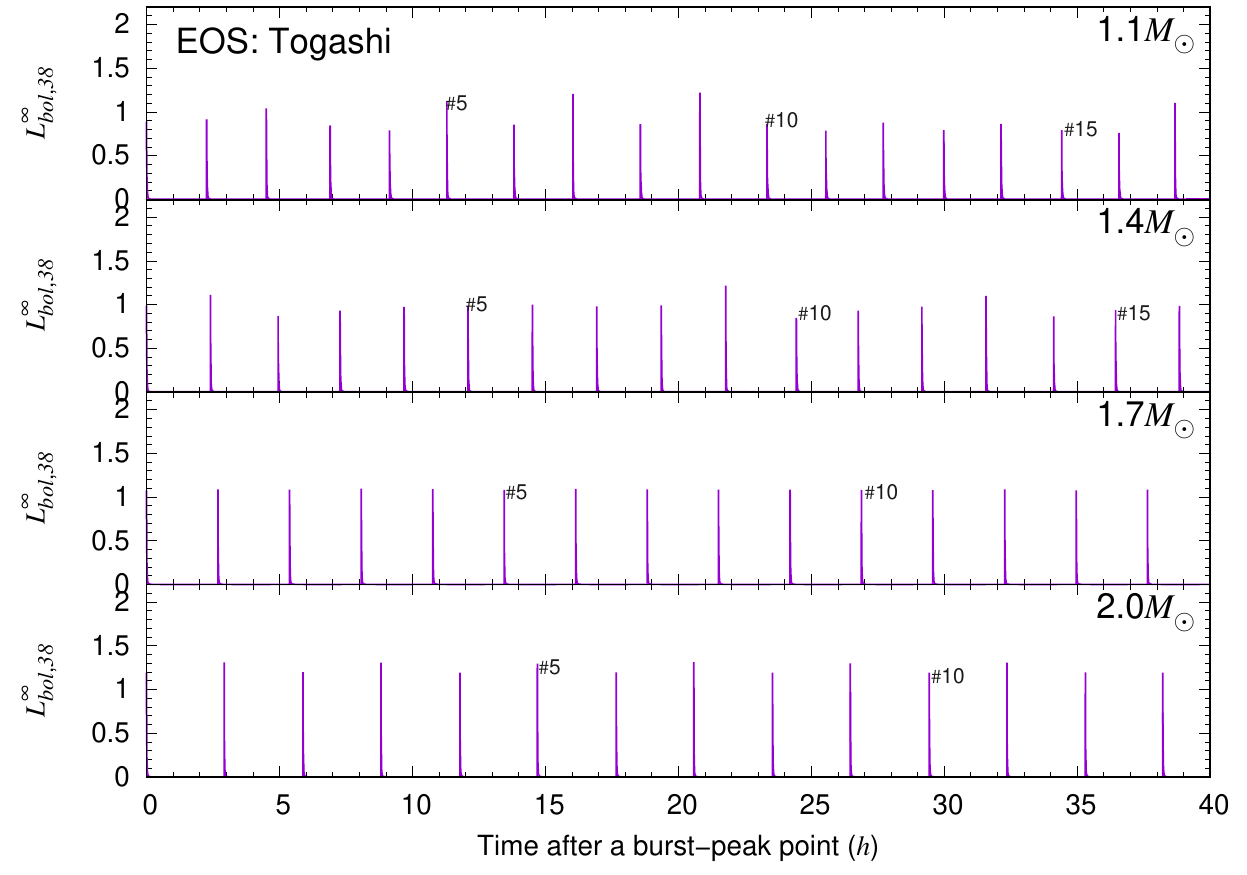}
    \caption{The luminosity of the burst sequence from $0$--$40~{\rm hr}$ for several $M_{\rm NS}$. Based on the Togashi EOS, we adopt $M_{\rm NS}=1.7M_{\odot}$, $\dot{M}_{-9}=2.5$, and $Z_{\rm CNO} = 0.01$.}
    \label{fig:Lall_Mass}
\end{figure*}

\begin{figure*}[h]
    \centering
    \includegraphics[width=0.8\linewidth]{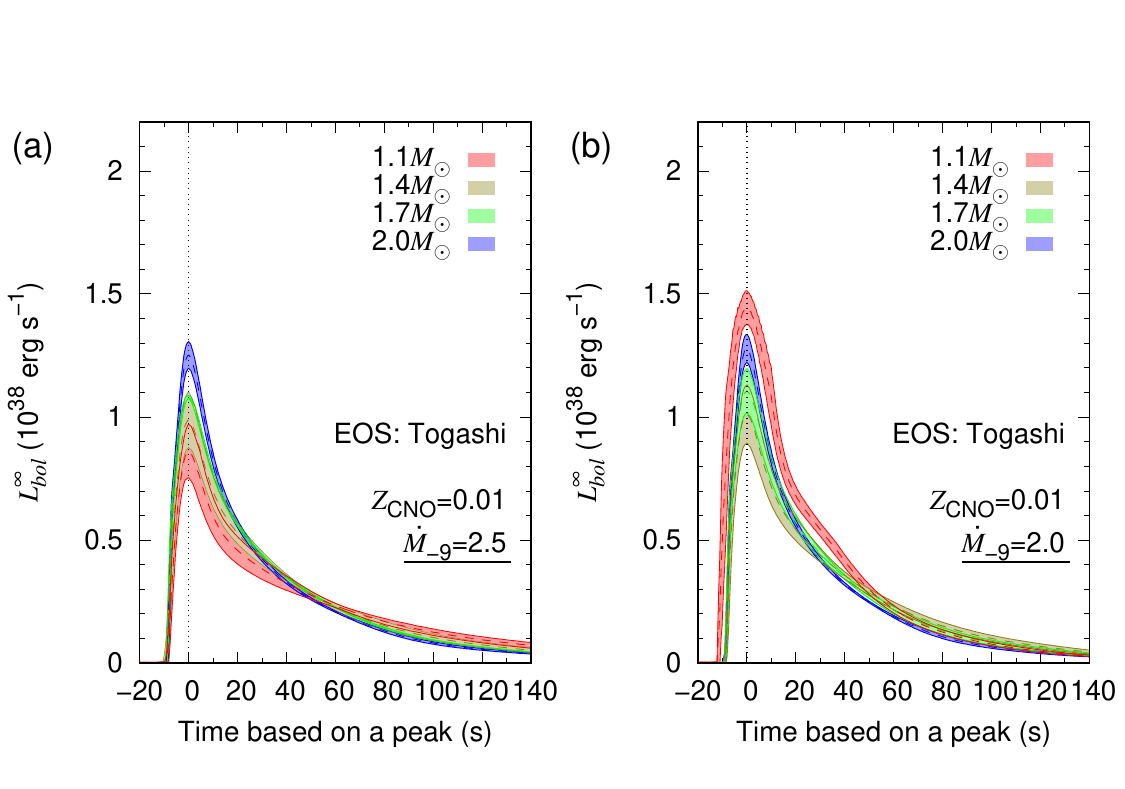}
    \caption{The light curves of the burst phase with the Togashi EOS for several $M_{\rm NS}$. Assuming $Z_{\rm CNO}=0.01$, we adopt (a) $\dot{M}_{-9}=2.0$ and (b) $\dot{M}_{-9}=2.5$.}
    \label{fig:Lb_Mass}
\end{figure*}

\subsection{The surface gravity and the ignition condition}
\label{sec:s_gravity}

We consider the physical conditions for the ignition of nuclear burning in the accreting layers on the surface of an NS. The gravitational acceleration at the NS surface $g_{\rm s}$ is useful to measure the strength of the surface gravity. The balance between $g_s$ and the pressure due to the mass accretion determines the ignition condition. We focus on $g_s$ and $P_{\rm ign}$ (the pressure at the time of nuclear ignition) as key quantities for the burst light curves.

We can derive a relationship between $g_s$ and $P_{\rm ign}$ based on a simplified one-zone burst condition, called the shell-flashed model, adopted in the our previous studies \citep{1999A&A...342..464K, 2004ApJ...603..242K}. This model approximately reproduces the structure of accretion layers during the flash, where the position of the NS shell is much lower than the pressure scale height due to the strong gravity \citep{1978PASJ...30..467S, 1981ApJ...247..267F}. The ignition pressure remains constant during the flash and is derived from the hydrostatic equation,
\begin{eqnarray}
P_{\rm ign} = g_{\rm s}\sigma, \ \ g_{\rm s} = \frac{GM_{\rm NS}}{R_{\rm NS}^2}\left(1 - \frac{2GM_{\rm NS}}{R_{\rm NS}c^2} \right)^{-1/2} ,
\label{eq:eq11}
\end{eqnarray}
where $\sigma$ is the column depth. The effect of general relativity is considered in $g_{\rm s}$; $G$ is the gravitational constant, and $c$ is the speed of light. In this one-zone framework, we assume all fuel in the shell is consumed in one burst event. Thus, using the energy generation density of nuclear burning ($Q_{\rm nuc}$), the burst energy $E^{1z}_b$ is expressed as
\begin{eqnarray}
E^{1z}_b = 4\pi R_{\rm NS}^2\sigma Q_{\rm nuc}\left(1+z_g\right) \label{eq:eq12}  \ .
\end{eqnarray}
A lower $R_{\rm NS}^2\left(1+z_g\right)\sigma$ with a compact NS causes a lower $E^{1z}_b$ when we consider only the effect of surface gravity ignoring other processes, e.g., the neutrino cooling.

We show the values of $\sigma$ for several $R_{\rm NS}$ and $M_{\rm NS}$ in Fig.~\ref{fig:FHM}. As expected from the monotonic feature of the NS mass--radius relations (Fig.~\ref{fig:mr}), $g_{\rm s}$ shows monotonically varies on the $R_{\rm NS}$--$M_{\rm NS}$ plane. As we see Eq.~(\ref{eq:eq11}) with a constant pressure condition, a more compact NS shows a higher $g_{\rm s}$ and lower $\sigma$. Therefore, a more compact NS has a less nuclear fuel (a smaller value of $\sigma$). This means the duration of nuclear burning becomes longer for more compact NSs, i.e., a higher luminosity in the tail but a lower luminosity near the peak. We expect, moreover, that a lower $\sigma$ takes more time than the accumulated matter from a companion to be ignited. The influence of $g_{\rm s}$ on multizone X-ray-burst models has already been discussed based on Eq.~(\ref{eq:eq11}) \citep{2020PTEP.2020c3E02D}, assuming the constant ignition pressure. In this work, however, we will show that such a previous discussion with only $g_{\rm s}$ and $\sigma$ is insufficient to explain the multizone X-ray-burst models due to the neutrino cooling effect. That is, we will see the influence of not only $g_{\rm s}$ but also $P_{\rm ign}$ on multizone X-ray-burst models in Section~\ref{sec:results}.

\subsection{The effect of neutrino cooling}
\label{sec:nu-cooling}

The temperature structure in steady state should depend on the NS mass or radius because the neutrino emissivity depends on the density. We present the mass dependence of the temperature structure in steady state in Fig.~\ref{fig:quies_mass}. As we see, if the mass is heavier, the temperature in steady state is much lower because the neutrino cooling is more enhanced owning to the central density being higher. By transporting the cooled heat from the core to the outside regions, we see that not only the inside but also the outside regions are cooled, which lowers the temperature around the ignition pressure $P_{\rm ign}\approx10^{22-23}~{\rm dyn~cm^{-2}}$.

This implies the possibility that neutrino cooling processes affect the burst light curve. Actually, several studies show this with using $Q_b$~\citep[e.g.,][]{2018ApJ...860..147M, 2020MNRAS.494.4576J}. Although their formulation, which covers only the accreted layer, has no choice but to give a physically groundless $Q_b$ as the boundary condition, our formulation enables us to calculate $Q_b$ without such an artificial parameter. $Q_b$ includes only the crustal heating energy, not the energy lost from neutrino cooling. So, we define the net base heat including the loss of neutrinos as $Q_{b+\nu}$, which is expressed as
\begin{eqnarray}
Q_{b+\nu} = \left(1.66\times10^{-35}~{\rm MeV~u^{-1}}\right)\frac{L_{\rm crust}}{\dot{M}_{-9}}~,\label{eq:9}
\end{eqnarray}
where $L_{\rm crust}$ is the luminosity on the NS crust in cgs units. $b+\nu$ means that the base heat includes not only crustal heating but also neutrino cooling unlike $Q_b$. In Table~\ref{tab:Q_beff}, we show the values of $Q_{b+\nu}$. The corresponding $Q_b$ values in our calculations are also listed in Appendix~\ref{sec:Qb}.

As seen in Table~\ref{tab:Q_beff}, the $Q_{b+\nu}$ value with higher-mass stars is lower due to lower temperature, which is caused by the neutrino cooling effect. Thus, $Q_b$ certainly depends on the mass (and radius). In the case of the crustal heating model by \cite{1990A&A...227..431H} and the slow neutrino cooling scenario assumed in this work, $Q_{b+\nu}$ is estimated to be around $0.5$--$0.7~{\rm MeV~u^{-1}}$. $Q_{b+\nu}$ also depends on the EOS, but compared with the mass dependency, the change of $Q_{b+\nu}$ or temperature structure due to the EOS is negligible within the slow neutrino cooling scenario. In this work, we do not consider the fast cooling scenario such as the nucleon direct Urca process, which works with high-symmetry-energy models such as the LS220 ($M_{\rm NS}\gtrsim 1.4M_{\odot}$) and TM1 EOSs \cite[e.g.,][]{2019PTEP.2019k3E01D}.

In Fig.~\ref{fig:Q-gs}, we show the correlation of $Q_{{\rm b} + \nu}$ and the surface gravitational acceleration $g_{{\rm s},14}$. They show anticorrelation for every EOS, where the effects of $Q_{{\rm b} + \nu}$, mainly neutrino cooling, becomes significant for higher $M_{\rm NS}$. The impacts depend on the relative value of $Q_{{\rm b} + \nu}$ and $g_{{\rm s},14}$. Stiff EOS (e.g., TM1), which has a larger NS radius, shows much large $Q_{{\rm b} + \nu}$ compared with $g_{{\rm s},14}$, while the softer EOS shows an opposite trend. Although the effect of gravity is dominated, we cannot ignore the effect of neutrino cooling. We discuss this in detail using our X-ray-burst models in the next session.

\section{Results}
\label{sec:results}

We calculate the thermal evolution of X-ray bursters from the initial conditions (described in Section~\ref{sec:init}), where energy generation by crustal heating balances the energy loss by neutrino cooling. At the early period of the outburst sequence, the energy generation tends to be higher than the burst events in the later phase due to the residual of the initial compositions in the accreted layer. This mechanism is known as ``compositional inertia'' \citep{1980ApJ...241..358T, 2004ApJS..151...75W}, which prevents a change in compositions in the accreted layer, converged at the early stage of the burst calculation. Therefore, we discard several decades of the burst with $ t \lesssim 2 \times 10^5~{\rm s} $ for all burst calculations. In all subsequent burst events, we select at least $15$ successive bursts for the following analysis.

\begin{figure*}[h]
    \centering
     \begin{minipage}{0.49\linewidth}
    \centering
    \includegraphics[width=0.9\linewidth]{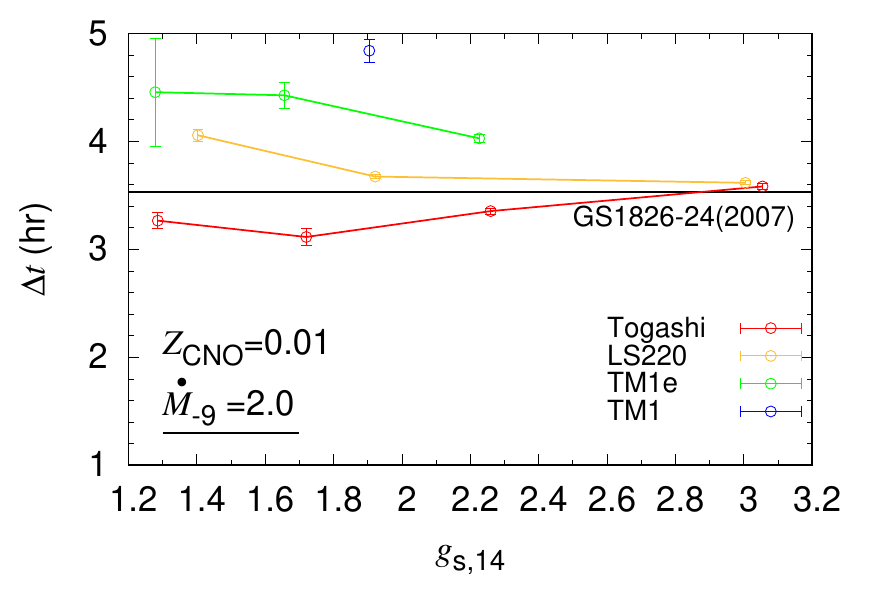}
     \end{minipage}
    \begin{minipage}{0.49\linewidth}
    \centering
    \includegraphics[width=0.9\linewidth]{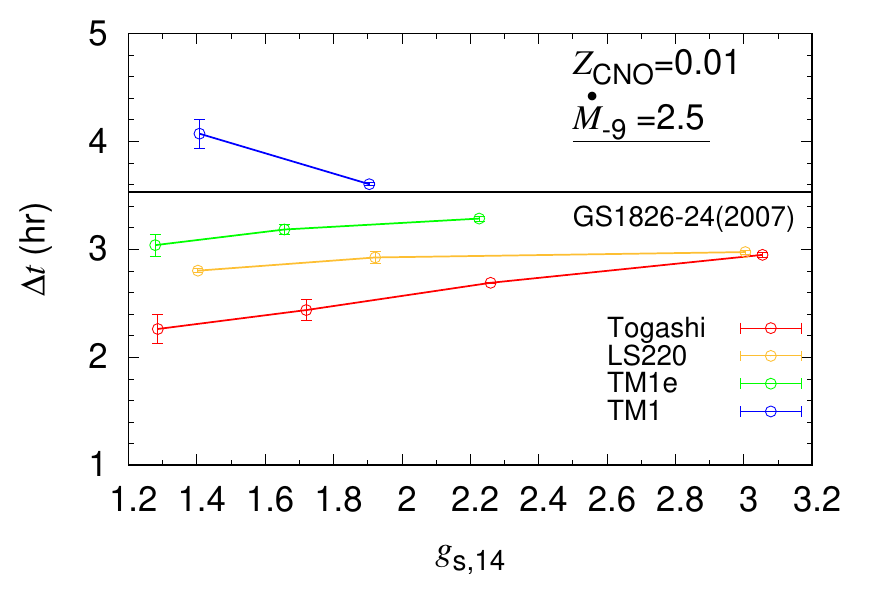}
    \end{minipage}
    \caption{The $\Delta t$ plotted by $g_{{\rm s},14}$ with different EOSs, i.e., Togashi (red), LS220 (orange), TM1e (green), and TM1 (blue). We adopt $Z_{\rm CNS} = 0.01$ with $\dot{M} = 2.0$ (left) and $\dot{M} = 2.5$ (right). The horizontal solid line corresponds to the observed value of GS~1826-24 \citep{2017PASA...34...19G}. Models with $L>L_{\rm edd}$ are omitted, where the hydrostatic equilibrium is invalid.}
    \label{fig:dt}
\end{figure*}

\begin{figure*}[h]
    \centering
     \begin{minipage}{0.49\linewidth}
    \centering
    \includegraphics[width=0.9\linewidth]{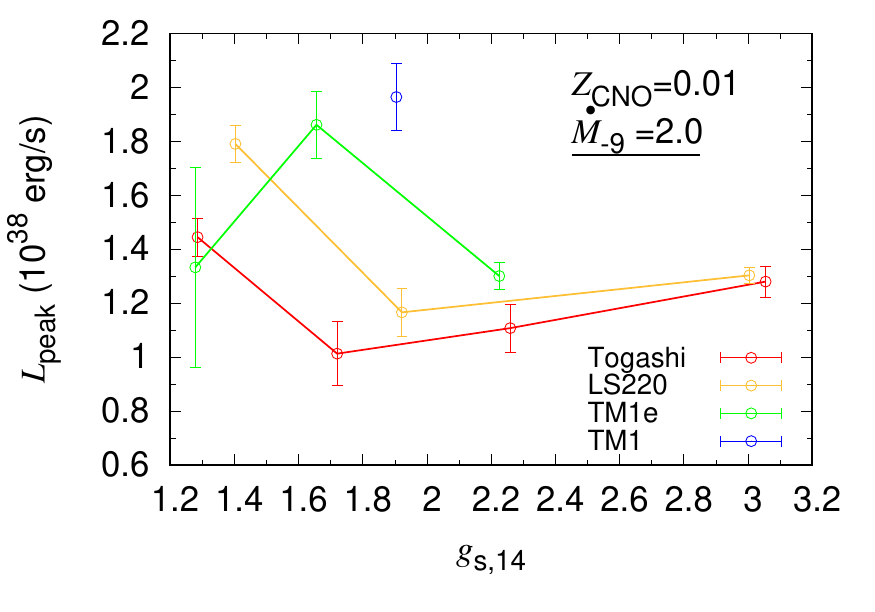}
     \end{minipage}
    \begin{minipage}{0.49\linewidth}
    \centering
    \includegraphics[width=0.9\linewidth]{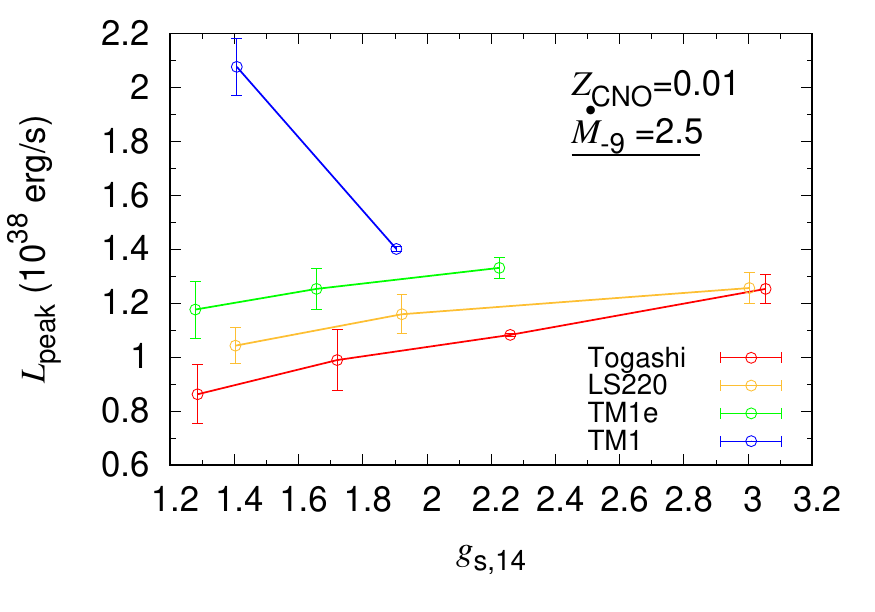}
    \end{minipage}
    \caption{Same as Fig.~\ref{fig:dt}, but for the peak luminosity $L_{\rm peak}$}
    \label{fig:Lpk}
\end{figure*}

\begin{figure*}[h]
    \centering
     \begin{minipage}{0.49\linewidth}
    \centering
    \includegraphics[width=0.9\linewidth]{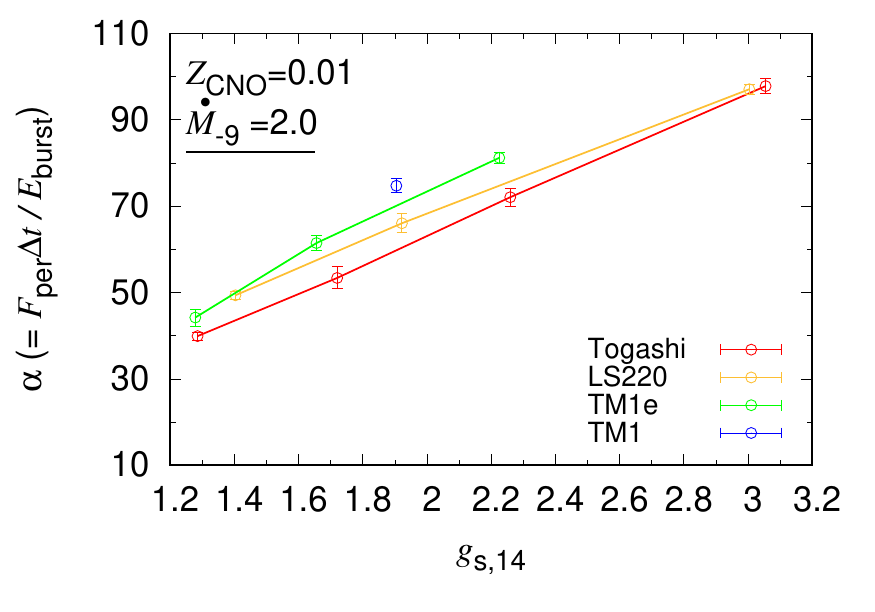}
     \end{minipage}
    \begin{minipage}{0.49\linewidth}
    \centering
    \includegraphics[width=0.9\linewidth]{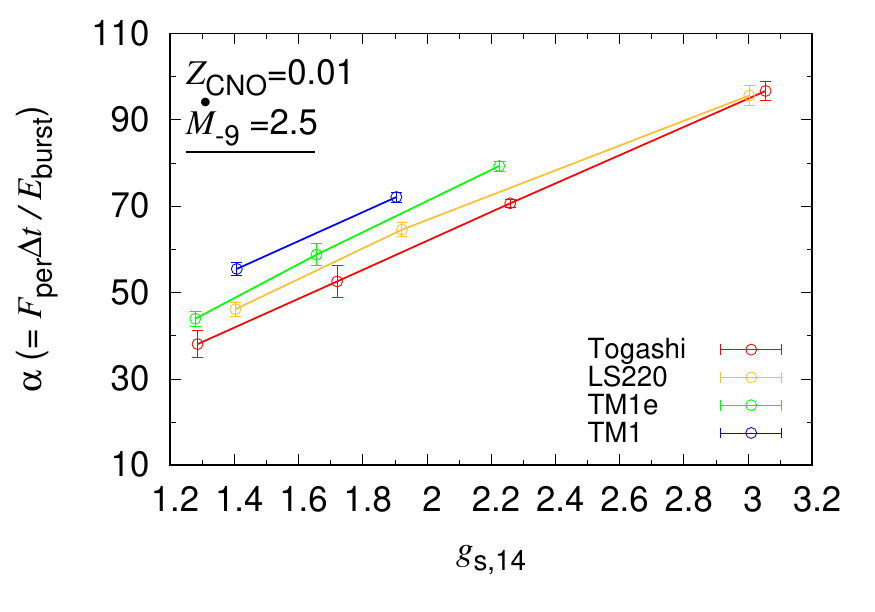}
    \end{minipage}
    \caption{Same as Fig.~\ref{fig:dt}, but for the burst strength $\alpha$}
    \label{fig:alpha}
\end{figure*}

\begin{figure*}[t]
    \centering
     \begin{minipage}{0.49\linewidth}
    \centering
    \includegraphics[width=0.9\linewidth]{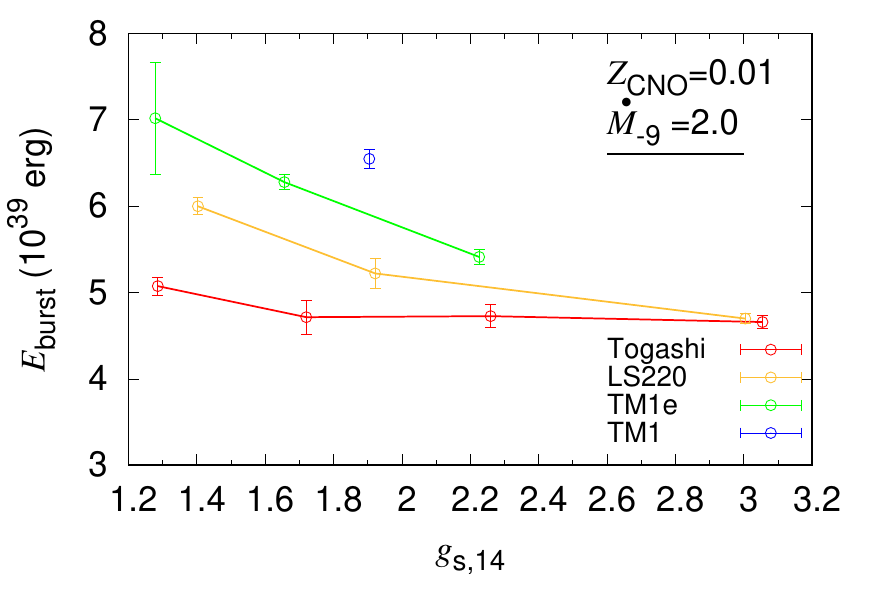}
     \end{minipage}
    \begin{minipage}{0.49\linewidth}
    \centering
    \includegraphics[width=0.9\linewidth]{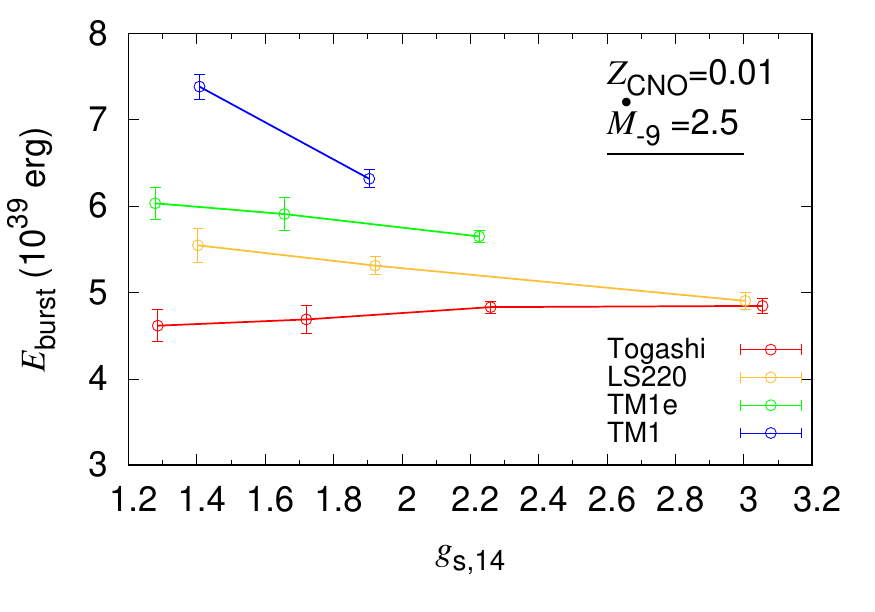}
    \end{minipage}
    \caption{Same as Fig.~\ref{fig:dt}, but for the total burst energy $E_{\rm burst}$}
    \label{fig:Eb}
\end{figure*}

\begin{figure}[ht]
    \centering
    \includegraphics[width=\linewidth]{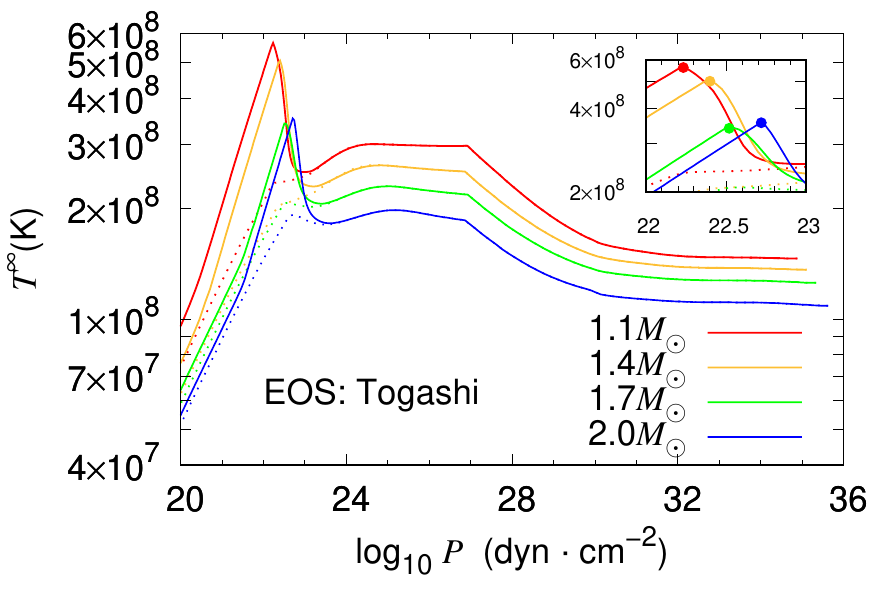}
     \centering
    \includegraphics[width=\linewidth]{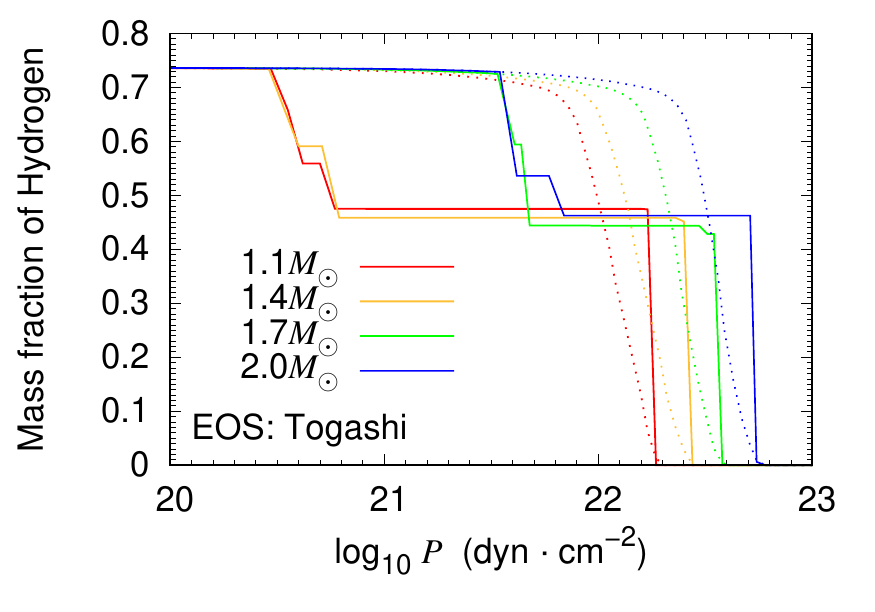}
    \caption{The structures of the redshifted temperature (upper) and the mass fraction of hydrogen (lower) plotted against the ignition pressure with $1.7~M_{\odot}$ stars. We adopted $1.7~M_{\odot}$, $\dot{M}_{-9}=2.5$, and $Z_{\rm CNO}=0.01$. For each model, the dotted line and the solid line that show the structures before and after ignition (at $t\simeq10^6~{\rm s}$), respectively.}
    \label{fig:strccomp_mass}
\end{figure}

\subsection{The Impacts of $M_{\rm NS}$ and EOS on X-Ray Bursts}

\subsubsection{X-Ray-Burst Light Curves}\label{sec:lc_eos}

In Figure \ref{fig:Lall_EOS}, we show the calculated light curves of the X-ray-burst models. The burst events of $M_{\rm NS}=1.7M_{\odot}$, $\dot{M}_{-9}=2.5$, and $Z_{\rm CNO} = 0.01$ with four different models are plotted. The time interval of each model is proportional to the NS radius, depending on the EOS. The NS radius becomes smaller in the following order: Togashi, LS220, TM1e, and TM1 (Fig.~\ref{fig:mr}), due to the softness of EOSs. The bursts with Togashi (soft EOS) show the smallest interval, while TM1 (stiff EOS) has the longest time interval. As the interval becomes larger, the peak luminosity also appears to be larger.

In Fig.~\ref{fig:Lb_EOS}, we compare the profiles of a typical burst light curve for different EOS burst models, with $N_{\rm NS} = 1.7$ and $M_{\rm CNO} = 0.01$. Fig.~\ref{fig:Lb_EOS}a shows cases of $\dot{M}_{-9}=2.5$. We find that the peak luminosity becomes higher for the larger NS radii, which is different from the relationship between the period and NS mass. In particular, burst models with the TM1 EOS ($M_{\rm NS}=2.0 M_{\odot}$) have a significantly high peak luminosity. For the cases with a lower mass accretion ($\dot{M}_{-9}=2.0$) in Fig.~\ref{fig:Lb_EOS}b, the change in the peak luminosity appears to follow the same trend. The peak luminosity reaches at $2\times 10^{38}~{\rm erg}~{\rm s}^{-1}$, near the Eddington luminosity, so that the TM1 case occurs in the photospheric radius expansion (PRE). On the other hand, the tail structure of the light curves, mainly determined by nuclear burning in the rp-process, appears to be independent on of the EOS with $\dot{M}_{-9}=2.0$ and $2.5$. Other timescales (i.e., the rising phase, transient to burst phase, and decay time) are also independent of the EOS.

We make a similar comparison in Fig.~\ref{fig:Lall_Mass} and \ref{fig:Lb_Mass} focusing on the difference in $M_{\rm NS}$. The profiles of the burst sequences for the $M_{\rm CNO} = 0.01$ and $\dot{M}_{-9}=2.5$ models are shown in Fig.~\ref{fig:Lall_Mass}. We find that the intervals change due to the mass of the NS. As $M_{\rm NS}$ increases, the time interval becomes larger (Fig.~\ref{fig:Lall_Mass}). The peak luminosity becomes higher (Fig.~\ref{fig:Lb_Mass}a), though their changes are smaller than the variation caused by the EOS (Fig.~\ref{fig:Lb_Mass}a). We should note that the $M_{\rm NS}$ and EOS are not determined independently, as shown in Fig.~\ref{fig:mr}. However, a soft EOS has a compact NS (a smaller radius for the same $M_{\rm NS}$), while a stiff EOS shows a less compact one (a larger radius for the same $M_{\rm NS}$). Based on the compactness of NSs, the trends in the mass dependence and EOS dependence are opposite. The time interval is higher with the larger-radius EOS (or the higher mass at the same radius) in Fig.~\ref{fig:Lall_EOS} and \ref{fig:Lall_Mass}, while it is lower with a higher mass. For the peak luminosity, a similar difference is seen.

To resolve the discrepancy between the mass and EOS dependence, the effect of neutrino emission via the NS mass may be a key issue. As we discussed in Section~\ref{sec:nu-cooling} and implied from the initial temperature structure (Fig.~\ref{fig:quies_mass}), neutrino emission inside the NS, which lowers the temperature itself, is involved with light curves in the latter case of mass dependence. Under the current assumption of the same neutrino processes in any EOS, we recognize that the difference in light curves between both tendencies is caused by the neutrino emission inside the NS. The effect of neutrino cooling may change the dependence of the burst parameters on the mass and EOS (Fig.~\ref{fig:quies_mass}). This is seen in Fig.~\ref{fig:Lb_Mass}, where the mass dependence of light curves is nonmonotonic and complicated when we change $\dot{M}_{-9}$. While the peak luminosity becomes higher with higher mass for $\dot{M}_{-9}=2.5$, conversely, the peak luminosity is lower for $\dot{M}_{-9}=2.0$. The inversion of the trend is not seen in \ref{fig:Lb_EOS}, which compares the EOS dependence between $\dot{M}_{-9}=2.5$ and $\dot{M}_{-9}=2.5$. It can be caused by neutrino cooling on the light curves, the impact of which can be larger than others. The details will be discussed in Section~\ref{sec:gs-qb}.

\subsubsection{The X-Ray-burst Model Parameters}
\label{sec:bp}

The light curves of X-ray bursts are characterized by few parameters such as the recurrence time $\Delta t$, the peak luminosity $L_{\rm peak}$, and the burst strength of nuclear burning $\alpha$ \citep[see, e.g.,][]{2017PASA...34...19G}. The $\alpha$ is defined by the ratio of the accretion energy to the burst energy, i.e.,
\begin{eqnarray}
    \alpha = \frac{z_g}{1+z_g}\dot{M}c^2\frac{\Delta t}{E_b},~\label{eq:eq10}
\end{eqnarray}
where $z_g$ is the gravitational redshift and $E_b$ is the total burst energy. To calculate $E_b$, we set the minimum value for the luminosity in the numerical integration. We calculate $E_b$ for $L > 0.25 L_{\rm acc}$, where $L_{\rm acc}$ is the accretion luminosity. In our calculations, $L_{\rm acc}$ is small enough to calculate $E_b$ without the loss of generality. This is because the peak luminosities in the present study are $\sim 10^{38} {\rm erg~s^{-1}}$, which is much higher than $L_{\rm acc} \sim 10^{35} {\rm erg~s^{-1}}$. The dependence of the integration range on $E_b$ may change with high $\dot{M}$ near the Eddington accretion rate $\sim 10^{-8} M_{\odot} {\rm yr}$ \citep[e.g.,][for millihertz QPO from H/He mixed nuclear burning]{2007ApJ...665.1311H}.

\begin{figure*}[htbp]
\centering
\begin{minipage}{0.45\linewidth}
\centering
\includegraphics[width=\linewidth]{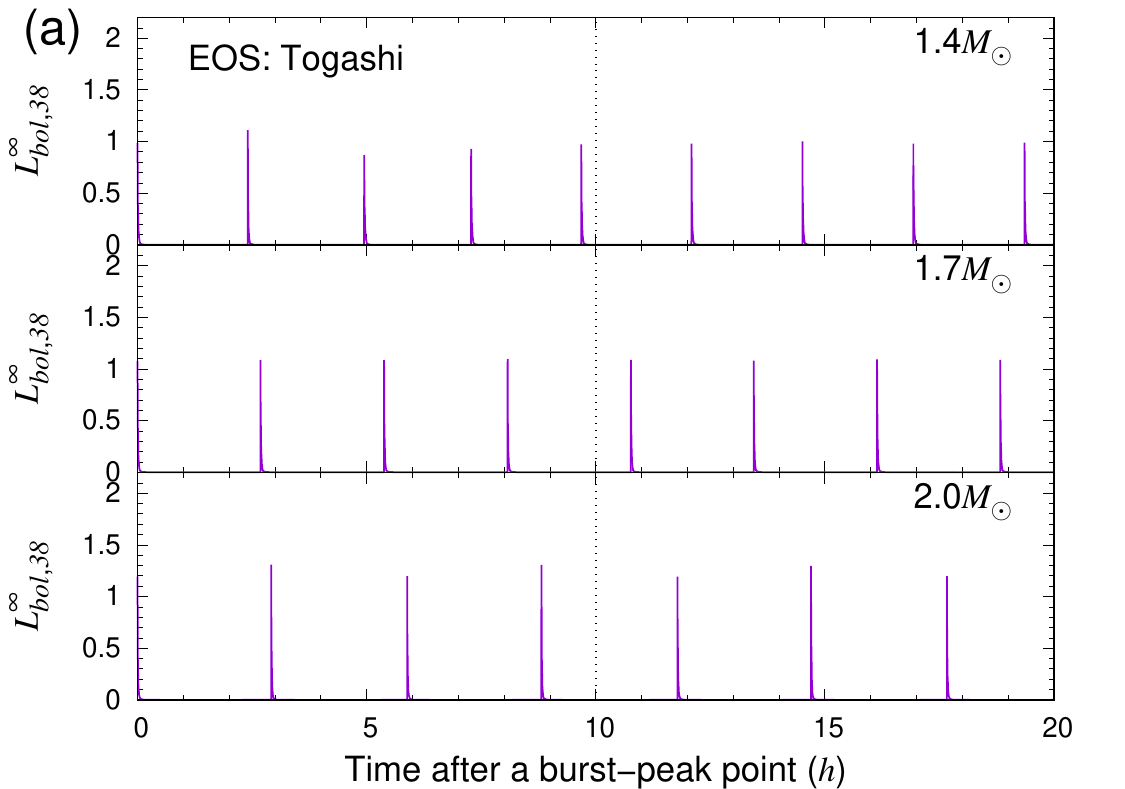}
\end{minipage}
\begin{minipage}{0.45\linewidth}
\centering
\includegraphics[width=\linewidth]{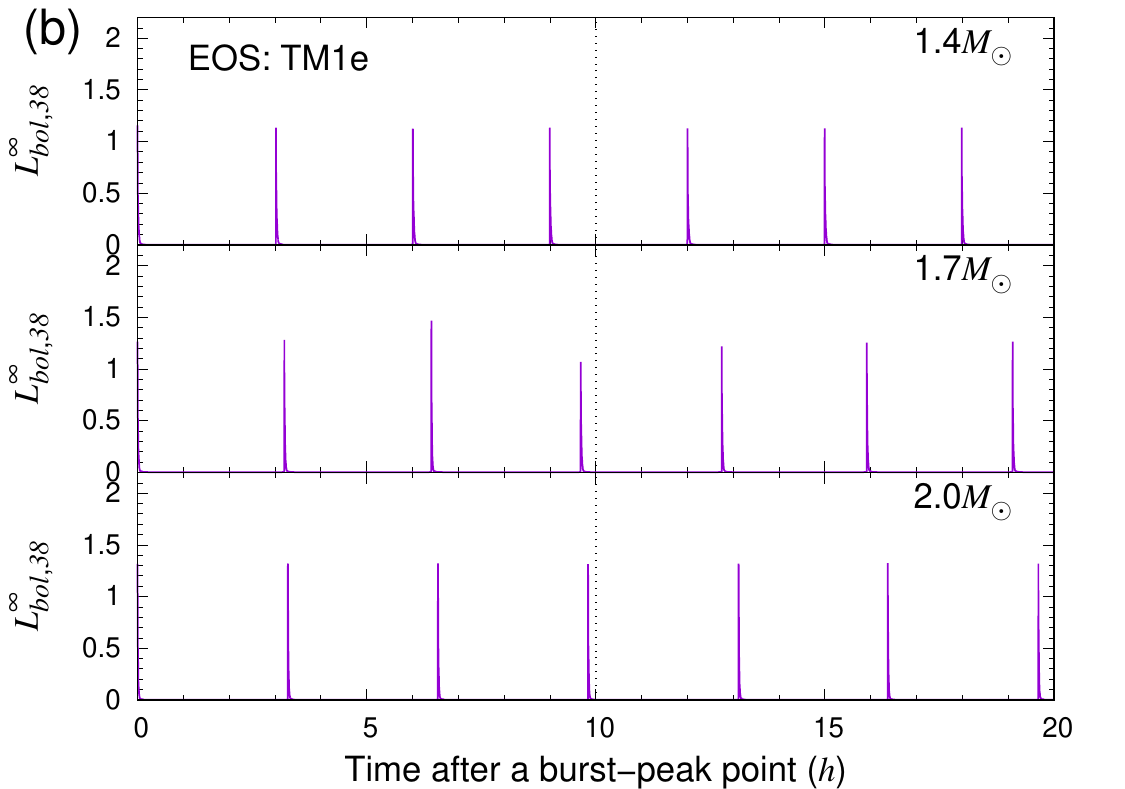}
\end{minipage}

\vspace*{0.5cm}
\begin{minipage}{0.45\linewidth}
\centering
\includegraphics[width=\linewidth]{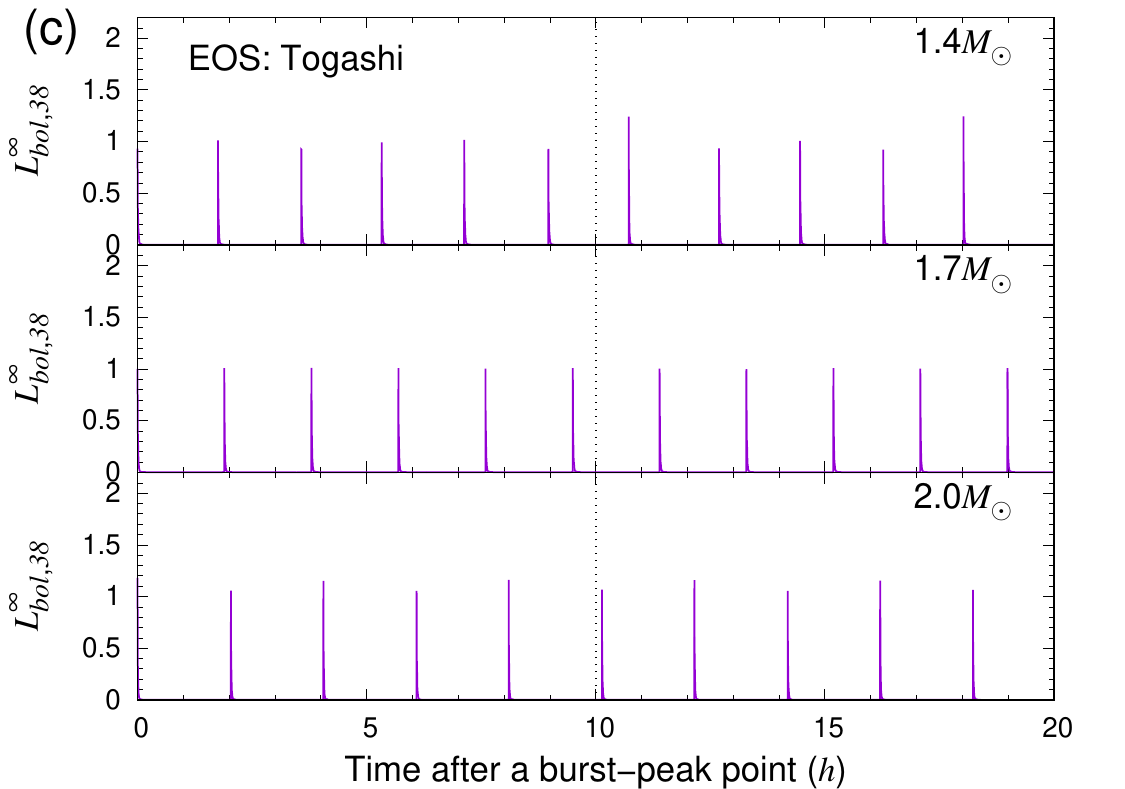}
\end{minipage}
\begin{minipage}{0.45\linewidth}
\centering
\includegraphics[width=\linewidth]{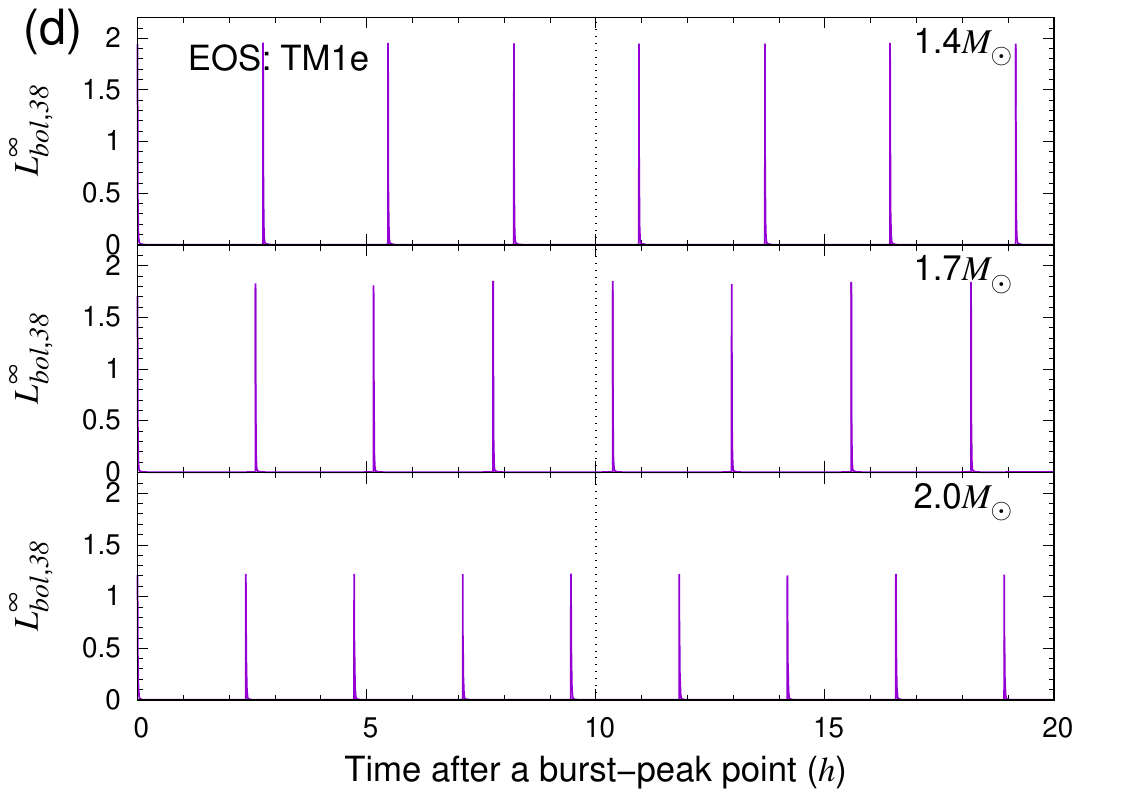}
\end{minipage}
\caption{The bolometric luminosity of the sequences of X-ray bursts over $20~{\rm hr}$. The results for $N_{\rm NS} = 1.4$, $1.7$, and $2.0 M_{\cdot}$ are shown in the each panel: (a) $(\dot{M}_{-9},Z_{\rm CNO}) = (2.5,0.01)$ with Togashi, (b) $(\dot{M}_{-9},Z_{\rm CNO}) = (2.5,0.01)$ with TM1e, (c) $(\dot{M}_{-9},Z_{\rm CNO}) = (3.0,0.02)$ with Togashi, and (d) $(\dot{M}_{-9},Z_{\rm CNO}) = (3.0,0.02)$ with TM1e. The time at $10~{\rm hr}$ is highlighted by the vertical dotted line.}
\label{fig:Lall}
\end{figure*}

\begin{figure*}[htbp]
    \centering
    \begin{minipage}{0.45\linewidth}
     \centering
    \includegraphics[width=\columnwidth]{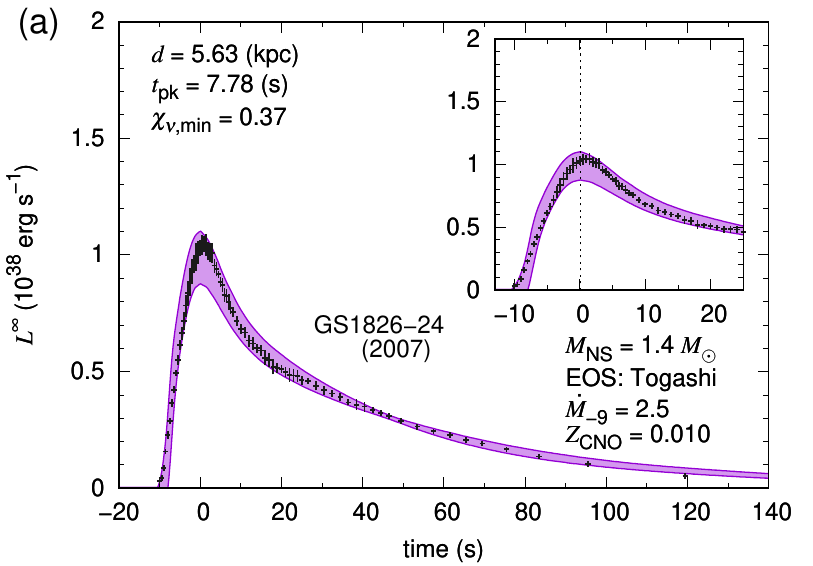}
     \centering
    \includegraphics[width=\columnwidth]{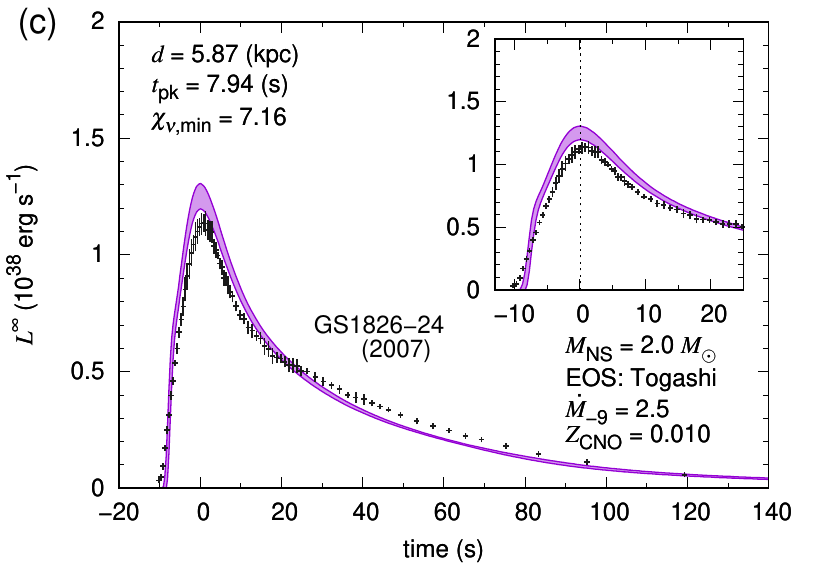}
    \end{minipage}
\begin{minipage}{0.45\linewidth}
\centering
    \includegraphics[width=\columnwidth]{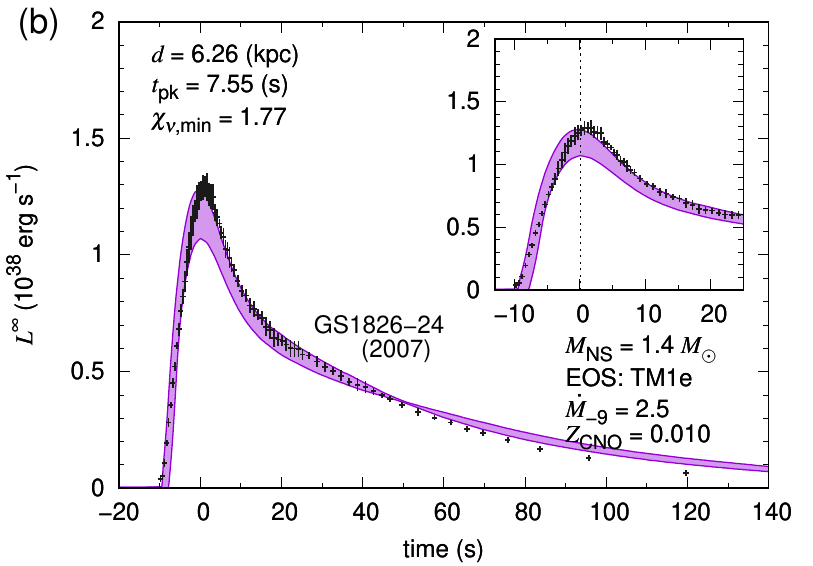}
    \includegraphics[width=\columnwidth]{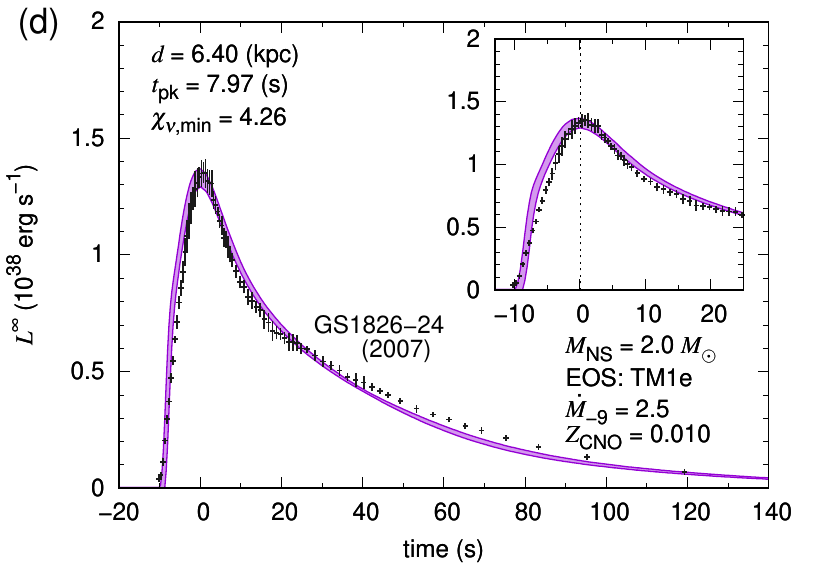}
    \end{minipage}
     \caption{Comparison of the calculated averaged burst light curves with observed ones of GS~1826-24 in 2007, assuming $\dot{M}_{-9}=2.5, Z_{\rm CNO}=0.01$: (a) Togashi with $M_{\rm NS}=1.4~M_{\odot}$, (b) Togashi with $2.0~M_{\odot}$, (c) TM1e with $M_{\rm NS}=1.4~M_{\odot}$, and (d) TM1e with $M_{\rm NS}=2.0~M_{\odot}$. We set $t=0~{\rm s}$ at the peak time of the light curves and plot within the $1 \sigma$ regions of many burst light curves. The distance including burst anisotropy $d$, the time after the peak $t_{\rm pk}$, and the minimum $\chi^2$ value $\chi_{\nu,{\rm min}}$ are shown in each panel.}
    \label{fig:GS}
\end{figure*}

Figures~\ref{fig:dt}, \ref{fig:Lpk}, \ref{fig:alpha} and \ref{fig:Eb} show $\Delta t$, $\alpha$, $L_{\rm peak}$, and $E_{\rm burst}$, respectively, for several $M_{\rm NS}$'s. As the NS mass is uniquely determined by the EOS with fixed burst parameters, we find a relationship between $M_{\rm NS}$ and the EOS. With few exceptions that depend on the EOS, the burst parameters (i.e., $\Delta t$, $L_{\rm peak}$, $\alpha$, and $E_{\rm burst}$) show a monotonic correlation with the NS mass. For the dependence on the NS mass, the behavior of $\Delta t$ and $L_{\rm peak}$ is hard to see because at least two different physical processes are involved with their change. However, $\alpha$ simply becomes larger with higher mass. This relationship of $\alpha$ is consistent with an EOS dependence in that the value is higher with more compact NSs.

In Fig.~\ref{fig:dt}, we compare the calculated $\Delta t$'s with the value of the X-ray binary of GS~1826-24 \citep{2017PASA...34...19G}. The burst models of $M_{\rm NS}=2.0M_{\odot}$, $Z_{\rm CNO}=0.01$, and $\dot{M}_{-9}=2.0$ with the Togashi and LS220 EOSs are consistent with observed values. Observational burst parameters (i.e., $\Delta t,L_{\rm peak}$, and $\alpha$) may be useful to constrain the EOS. Considering the uncertainty due to $M_{\rm NS}$, $\alpha$ provides the strictest restriction on the information of interior NS in many parameters related to the X-ray burst, though the properties of burst light curves are still sensitive to other parameters, especially to $\dot{M}_{-9}$ and $Z_{\rm CNO}$~\citep[e.g.,][]{2016ApJ...819...46L}.

\subsection{The Effects of $g_s$ and $Q_{b+\nu}$ on the X-Ray-burst Models}\label{sec:gs-qb}

We discuss physical reasons for the mass dependence of burst light curves. Assuming that the NS mass tends to have an anticorrelation with the NS radius in especially high-mass regions, though there are several exceptions, $E_b$ should be higher in high-mass models owning to the strong surface gravity. However, the mass dependence (shown in Fig.~\ref{fig:Eb}) does not match this tendency and therefore is affected by another effect. It is presumed to be the effect of neutrino emission to decrease the temperature as shown in the initial burst models in Fig.~\ref{fig:quies_mass}. This effect is characterized by the effective base heat $Q_{b+\nu}$ as seen in Fig.~\ref{sec:init}.

The $Q_{b+\nu}$ effect can be clearly seen in the mass dependence of the temperature structure and hydrogen mass fraction in Fig.~\ref{fig:strccomp_mass}. These figures show that the ignition pressure is higher with high-mass models. This cannot be explained by the $g_s$ effect, but by $Q_{b+\nu}$ effect, which plays a role to lower the overall temperature with higher-mass regions (Table~\ref{tab:Q_beff}). The pressure at peak temperature can judge whether $g_s$ or $Q_{b+\nu}$ effect is stronger, depending on the burst model.

For the mass dependence of the burst models, the effect of $g_s$ and $Q_{b+\nu}$ can be see simultaneously. In Fig.~\ref{fig:Lall}, the $\dot{M}_{-9}=2.5$ and $Z_{\rm CNO}=0.01$ case, both EOS models (Togashi and TM1e) have fewer burst events or higher $\Delta t$ with higher-mass regions. For these burst models, therefore, the $Q_{b+\nu}$ effect is larger than the $g_s$ effect. In Fig.~\ref{fig:Lall}, the $\dot{M}_{-9}=3.0$ and $Z_{\rm CNO}=0.02$ case, however, the number of burst sequences with the Togashi EOS is decreased with higher-mass regions and $\Delta t$ is higher, which is opposite to the case in the TM1e EOS. This implies that in the $\dot{M}_{-9}=3.0$ and $Z_{\rm CNO}=0.02$ case, the $Q_{b+\nu}$ effect is larger than the $g_s$ effect with the Togashi EOS while it is lower with the TM1e EOS. Thus, by changing $\dot{M}_{-9}$ and $Z_{\rm CNO}$, mass dependence of $\Delta t$ could changed qualitatively.

For the overall mass dependence of $\Delta t$, let us look at Fig.~\ref{fig:dt}. With the Togashi EOS, for example, $\Delta t$ basically has a positive correlation with mass, which means that the $Q_{b+\nu}$ effect is stronger than the $g_s$ effect. For other EOSs, however, that tendency does not always remain, such as with the TM1e EOS case with $\dot{M}_{-9}=3.0$ and $Z_{\rm CNO}=0.02$. Assuming that the quantitatively same $Q_{b+\nu}$ effect works with any EOS-like setting in this work, the $g_s$ effect appears more easily with the softer or higher-symmetry-energy EOS in high-density regions, such as the TM1 and TM1e EOSs. This is because the $g_s$ effect is higher with a larger-radius EOS as shown in Fig.~\ref{fig:FHM}. 

In summary, the effect of the surface gravity is lower in higher-mass models, and this makes $\Delta t$ longer because more fuel needs to be ignited. Meanwhile, the $Q_{b+\nu}$ effect is higher, and this leads to shorter $\Delta t$. This is the reason why the effects of $g_s$ and $Q_{b+\nu}$ (Fig.~\ref{fig:Q-gs}) show opposite dependence on $\Delta t$. This might be useful in specifying not only the NS structure but also the heating or cooling processes inside NSs via the $Q_{b+\nu}$ effect in the future, though elucidating this is hard in that burst light curves are sensitive to other input parameters such as $\dot{M}_{-9}$ and $Z_{\rm CNO}$. 

For $L_{\rm peak}$, the tendency of the EOS and mass is similar to that of $\Delta t$. Because $L_{\rm peak}$ is highly related to how much hydrogen is burnt, a large amount of hydrogen is not burnt, as seen in Fig.~\ref{fig:strccomp_mass}. This dramatically affects the efficiency of the first nuclear burning in X-ray bursts, the triple-$\alpha$ reaction. Because the parameter dependence of the amount of fuel is unclear within the multizone framework, the dependence on the EOS and mass of $L_{\rm peak}$ is more unclear than that of $\Delta t$.

Despite the complicated mass dependence of $\Delta t$, $\alpha$ has a positive correlation with mass and an anticorrelation with the radius. Because $\Delta t$ and $E_{\rm burst}$ are higher with a larger-radius EOS, it is hard to reveal the reason, but due to the $Q_{b+\nu}$ effect, the neutrino flux increases in higher-mass regions, and this makes the persistent flux $F_{\rm per}$ higher. This is because, in the persistent term, the crustal heating process and neutrino cooling process are dominant for the thermal evolution of accreting NSs. Their strength is not that different from nuclear burning. Moreover, as $F_{\rm per}$ depends on $z_g/(1 + z_g)$, where $z_g$ is the gravitational redshift, as shown in Eq.~(\ref{eq:eq10}), the gravitational effect seems to be important for the mass dependence of $\alpha$. These are two of the reasons for the positive correlation between $\alpha$ and mass.  

\subsection{Application to the Clocked Burster GS1826-24}\label{sec:obs}

To examine the validity of the burst models, we compare the theoretical burst light curves and observations. For such an observation, we take the observed light curve of a clocked burster event with GS 1826-24 in 2007, whose burst number was 10 times. We show the EOS (Togashi, TM1e) and mass ($M_{\rm NS}=1.4,2.0~M_{\odot}$) dependence on the burst light curve in Fig.~\ref{fig:GS} with $\dot{M}_{-9}=2.5$ and $Z_{\rm CNO}=0.01$. If the EOS is changed from Togashi to TM1e, the peak luminosity becomes higher due to the $g_s$ effect. If the mass is changed from $M_{\rm NS}=1.4~M_{\odot}$ to $2.0~M_{\odot}$, however, the qualitative change is different for the EOS; the peak luminosity is lower for the Togashi EOS while higher for the TM1e EOS. This is explained by the $g_s$ and $Q_{b+\nu}$ effects. In Fig.~\ref{fig:GS}, the Togashi EOS with $M_{\rm NS}=1.4~M_{\odot}$ seems to be favored, which is plausible compared with other models. In our burst models, the larger-radius EOS has a higher luminosity or causes PRE, such as the TM1 EOS. However, it is known that all bursts of GS 1826-24 in 2007 did not show PRE~\citep[][]{2017PASA...34...19G}. Finally, for a  large-radius EOS such as TM1, the peak luminosity is too high to explain the observations due to a larger $g_s$ effect. Hence, observations of clocked bursters can possibly constrain the NS EOS. Things similar to the EOS dependence can apply to the mass dependence. In Fig.~\ref{fig:GS}, high-mass models seem to be inconsistent with the observations because the peak luminosity is lower than the observed light curves due to the larger $Q_{b+\nu}$ effect. Thus, the mass of clocked bursters might be constrained by the observations. Although burst models are very sensitive to the accretion rate and composition of the companion star, the $g_s$ and $Q_{b+\nu}$ effects cannot be ignored in consistent burst modeling of bursters.

In this work, we could not find consistent models with both $\Delta t$ and the shape of the burst light curve of GS 1826-24 in 2007 at the same time. For example, we show the well-fitted burst model: Togashi EOS, $M_{\rm NS}=1.4~M_{\odot}$, $\dot{M}_{-9}=2.5$ and $Z_{\rm CNO}=0.01$. This burst model shows a shorter recurrence time $\Delta t = 2.44\pm0.10~{\rm h}$ compared with the observed one of $\sim3.53~{\rm h}$. Thus, it is not easy to find consistent burst models with  GS 1826-24 in 2007, but such a burst model can be created by changing the mass and EOS. Actually, we have previously presented the consistent burst model of ``L2n20Z1" in \citep[][see also Figures.~5 and 6 and Table 2 in this reference]{2020PTEP.2020c3E02D}, where the parameters are for the LS220 EOS and $M_{\rm NS}=1.58M_{\odot}$, $\dot{M}_{-9}=2.0$, and $Z_{\rm CNO}=0.01$. The effective base heat is calculated to be $Q_{b+\nu}=0.67~{\rm MeV~u^{-1}}$ ($Q_b=0.35~{\rm MeV~u^{-1}}$).

If the radius is large, such as in the TM1 EOS, the peak luminosity tends to be too high to explain the shape of burst light curves of GS 1826-24 due to the $g_s^{-1}$ effect. PRE occurs in several models, which are inconsistent with the observations of GS 1826-24 in 2007. Hence, a large-radius EOS does not seem to be preferred from the observations of GS 1826-24, and this is consistent with other observational constraints. Although the EOS and mass dependences of burst light curves are complicated due to the competition between the $g_s$ and $Q_{b+\nu}$ effects as shown above, if more information on the companion is available in the future, observational bursters can play a significant role in constraining the NS EOS and mass.

\section{Summary and conclusions}\label{sec:con}

In this study, we investigated theoretical models of X-ray bursts using a 1D general-relativistic evolution code with detailed microphysics. We performed a set of X-ray-bursts models, the light curves of which are compared with observations. We found the microphysics of NSs (i.e., the EOS and cooling and heating) significantly affected the theoretical prediction of the X-ray light curves. The results are summarized as follows.

\begin{itemize}
    \item The burst parameters characterizing the X-ray-burst light curves depend on the microphysics of the NS interior (Fig.~\ref{fig:dt}--\ref{fig:Eb}). The recurrence time ($\Delta t$) and the peak luminosity ($L_{\rm peak}$) have a positive correlation with the radius of the NS. The $\alpha$ has a monotonic correlation with the surface gravity. However, the uncertainty of the EOS affects the NS mass and radius relation and can have significant impacts on the burst parameters.

    \item The NS cooling may vary the burst light curves (i.e., burst parameters) even in the slow cooling scenario. If the mass is higher, $\Delta t$ and $L_{\rm peak}$ are higher owning to the neutrino cooling. As the temperature in the accreted layer is reduced by NS cooling, the required ignition pressure becomes higher. 

    \item Even considering NS cooling, $\alpha$ shows a strong correlation to the surface gravity (Fig.~\ref{fig:alpha}). Thus, among the burst parameters, the $\alpha$ may be the primary parameter to constrain the NS mass and radius.

    \item We constrained the mass and EOS of the NS GS~1826-24 by comparing observations of X-ray-dburst events in 2017. Generally, the models with a stiffer EOS, resulting in a larger NS radius, are ruled out.

\end{itemize}

We showed that the microphysics of the NS, such as the EOS and neutrino cooling, is important for theoretical X-ray-burst models. On the other hand, many previous works, even based on multizone X-ray burst models, treat $Q_b$ as an artificial parameter instead of considering the neutrino cooling effect. In this study, the effective base heat $Q_{b+\nu}$ varies with the mass and affects the burst light curves and parameters. Thus, our study implies that the base heat should not be treated as an artificial parameter for more realistic burst models.

As an observational reference, we took the clocked burster events of GS~1826-24 in 2007. We compared our burst models with the observed light curves. As a result, we attempted to constrain the NS EOS and the mass of GS~1826-24. The X-ray binary parameters (e.g., the accretion rate and chemical composition of accreted matter) may change our constraints on the EOS and the NS mass. However, we can develop a deeper understanding if we have more information on the accreted matter and accretion rate.

For the neutrino energy loss included in $Q_{b+\nu}$, we assume the slow cooling processes. However, this assumption may be insufficient in modern cooling theory \citep[e.g.,][]{2013arXiv1302.6626P}. One of the most important physics to describe the neutrino emissivity is the superfluidity of nucleons. Once the temperature in the NS layers drops below the critical temperature, the nuclear matter transits to a superfluid state, and it proceeds via the following two mechanisms of neutrino emissivity: nucleon superfluidity suppresses conventional neutrino emissions and opens an additional cooling channel, which we call the pair breaking formation process. The efficiency of these effects depends on the superfluid models. Various types of the superfluid phase can appear in a single NS simultaneously~\citep[e.g.,][]{2013ApJ...765....1N}. Pulsar glitches correspond to the neutron singlet pairing superfluidity, which occurs in the crust of the NS, and the observation of Cassiopeia A may require the triplet pairing of neutrons in the core~\citep{2011PhRvL.106h1101P, 2011MNRAS.412L.108S}. Hence, X-ray-burst calculations with neutrino cooling including the effect of nucleon superfluidity would be necessary to make more realistic burst models.

Also, the fast neutrino cooling process is required by some observations of cold NSs. The fast cooling process could occur under some specific conditions. Its emissivity is much larger than that of slow cooling processes, and it affects the thermal evolution of the NS, including X-ray bursts. The fast cooling process that likely occurs in NSs is the nucleon direct Urca process. It works in the region where the proton fraction exceeds the threshold by $1/9$. In our models, the nucleon direct Urca process occurs for the LS220 and TM1 EOSs due to their high symmetry energy, though we turned it off throughout this paper. Other fast cooling processes with exotic states of matter could also occur \citep[e.g.,][]{2004ARA&A..42..169Y}. Even within the slow cooling processes, the burst light curves and parameters are affected in this paper. Hence, compared with slow cooling processes, fast cooling processes could significantly change the $Q_{b+\nu}$ value and finally affect the burst light curves and parameters.

Modifying the heating processes as well as the cooling processes inside the NSs is important to describe the burst light curve. One important heating process is crustal heating. Despite some recent work to investigate the heating rate \citep{2018ApJ...859...62L, 2018A&A...620A.105F, 2021arXiv210501991S}, the amount of heat released from total crustal heating processes has large uncertainties (typically $1$--$2~{\rm MeV~u^{-1}}$). Moreover, the efficiency in the heat transport from the source to the NS surface of accreted layers depends on the NS model. These uncertainties in the heating process definitely affect burst light curves. Such a detail of the heating and cooling effects on the X-ray burst is left for our future work.
\\

 This project was financially supported by JSPS KAKENHI (19H00693, 20H05648, 21H01087) and a RIKEN pioneering project ``Evolution of Matter in the Universe (r-EMU)''. N.N. was supported by Incentive Research Project (IRP) at RIKEN. Parts of the computations in this study were carried out on computer facilities at CfCA, National Astronomical Observatory of Japan and at YITP, Kyoto University.

\appendix

\section{Artificial parameter $Q_b$ without neutrino cooling}
\label{sec:Qb}

Our calculated base heat, $Q_{b+\nu}$, includes the neutrino emission inside the NS core, but many studies for examining the thermal evolution of accreting NSs distinguish such a neutrino cooling effect from the base heat because the neutrino emission does not contribute to the heating in cold NSs. In their formulation, the base heat is treated as the parameter $Q_b$, not $Q_{b+\nu}$. So, we also show calculated values of $Q_b$ as follows:
\begin{eqnarray}
Q_{b} = \left(1.66\times10^{-35}~{\rm MeV~u^{-1}}\right)\frac{L_{\rm crust}-L_{\nu}}{\dot{M}_{-9}}~,\label{eq:app}
\end{eqnarray}
where $L_{\nu}$ is the neutrino luminosity on the NS crust in cgs units. The results are shown in Table~\ref{tab:Q_b}, assuming the crustal heating processes modeled by \cite{1990A&A...227..431H}. For all models, our $Q_b$ value is around $0.3$--$0.4$~${\rm MeV~u^{-1}}$, which is consistent with the constraint of $Q_b<0.5~{\rm MeV~u^{-1}}$ implied by \cite{2018ApJ...860..147M} and \cite{2020MNRAS.494.4576J}. The dominant component of $Q_b$ is the crustal heating process, which is treated to be in proportion to $\dot{M}_{-9}$. Therefore, even if $\dot{M}_{-9}$ is changed in our adopted regions, our $Q_b$ values are not affected at all. Compared with $Q_{b+\nu}$, $Q_b$ is clearly lower, and this means that the neutrino cooling effect should be considered even if an artificial parameter $Q_b$ is introduced within the hydrodynamic simulation.

\begin{table}[h]
    \centering
     \caption{Calculated $Q_{b}~[{\rm MeV~u^{-1}}]$ values with different EOSs and Masses.}
     \hspace*{-2cm}
    \begin{tabular}{ccccc}
    \hline\hline
    $M_{\rm NS}$     &  Togashi & LS220 & TM1e & TM1 \\
         \hline
    $1.1 M_{\odot}$     & 0.39 & 0.39 & 0.39 & 0.40 \\
    $1.4 M_{\odot}$     & 0.35 & 0.37 & 0.37 & 0.37 \\
    $1.7 M_{\odot}$     & 0.34 & 0.34 & 0.36 & 0.36 \\
    $2.0 M_{\odot}$     & 0.30 & 0.31 & 0.33 & 0.34 \\
    \hline
    \end{tabular}\\
         {\textbf{Note.} $Q_{b}~[{\rm MeV~u^{-1}}]$ includes the components of crustal heating, photon cooling, and the effect of gravitational shrinking.}
    \label{tab:Q_b}
\end{table}

\section{Physical parameters of burst models}

In Table \ref{tab:data}, we summarize our burst models. If you choose mass, radius, initial metallicity, and accretion rate assuming $X/Y=2.9$, you can obtain several burst parameters: burst strength $\alpha$; burst duration $\tau$, which is defined to be the time after the peak at half the value of the peak luminosity; recurrence time $\Delta t$; total burst energy $E_{\rm burst}$; peak luminosity $L_{\rm peak}$; rise time from transience to peak point $t_{\rm rise}$; and e-folding time after peak point $\tau_e$. For all output parameters, each error with $1\sigma$ regions is also shown. We only show the data with the Togashi and TM1e EOSs and $M=1.4, 2.0~M_{\odot}$ in Table~\ref{tab:data}. The numerical data are available in a supplemental table.

\begin{table*}[ht]
		\caption{Physical quantities of some burst models. Uncertainties of output values indicate the standard deviation 1$\sigma$. Some burst models with PRE are not followed.}
		\label{tab:data}
		\begin{center}
		\hspace*{-2.0cm}
			\scalebox{0.8}{
			\begin{tabular}{cccccccccccc}
			\vspace{-1.0cm} \\ \hline\hline
			EOS & $M_{\rm NS} [M_{\odot}]$ & $R_{\rm NS} [{\rm km}]$ &  $Z_{\rm CNO}$ &
			$\dot{M}_{-9}$ &
			$\alpha$ &
			$\tau [{\rm s}]$ &
			$ \Delta t [{\rm h}]$ &
			$E_{\rm burst} [10^{39} {\rm erg}]$ &
			$L_{\rm peak} [10^{38} {\rm erg~s^{-1}}]$ & $t_{\rm rise} [{\rm s}]$ &
		$\tau_e [{\rm s}]$\\
				\hline
Togashi & 1.4& 11.601& 0.005& 2.0&  52.6$\pm$ 3.1&  52.3$\pm$ 5.1&  3.39$\pm$0.13&  5.21$\pm$0.20&  1.01$\pm$0.14&  5.15$\pm$0.43&  39.0$\pm$ 3.5
\\ Togashi & 1.4& 11.601& 0.005& 2.5&  51.4$\pm$ 1.1&  55.6$\pm$ 1.6&  2.61$\pm$0.03&  5.13$\pm$0.10&  0.92$\pm$0.03&  4.99$\pm$0.35&  40.7$\pm$ 2.1
\\ Togashi & 1.4& 11.601& 0.005& 3.0&  51.3$\pm$ 2.8&  58.0$\pm$ 2.2&  1.94$\pm$0.07&  4.58$\pm$0.13&  0.79$\pm$0.05&  4.83$\pm$0.37&  42.4$\pm$ 2.8
\\ Togashi & 1.4& 11.601& 0.005& 4.0&  51.2$\pm$ 2.6&  60.6$\pm$ 2.4&  1.57$\pm$0.03&  4.95$\pm$0.17&  0.82$\pm$0.06&  5.29$\pm$0.46&  44.5$\pm$ 2.5
\\ Togashi & 1.4& 11.601& 0.01 & 2.0&  53.4$\pm$ 2.6&  46.9$\pm$ 3.3&  3.11$\pm$0.08&  4.71$\pm$0.20&  1.01$\pm$0.12&  4.81$\pm$0.37&  36.1$\pm$ 2.9
\\ Togashi & 1.4& 11.601& 0.01 & 2.5&  52.6$\pm$ 3.8&  47.8$\pm$ 3.9&  2.44$\pm$0.10&  4.69$\pm$0.16&  0.99$\pm$0.11&  4.90$\pm$0.44&  35.7$\pm$ 3.3
\\ Togashi & 1.4& 11.601& 0.01 & 3.0&  51.4$\pm$ 1.3&  52.7$\pm$ 1.7&  1.97$\pm$0.02&  4.63$\pm$0.09&  0.88$\pm$0.04&  4.64$\pm$0.33&  38.6$\pm$ 2.1
\\ Togashi & 1.4& 11.601& 0.01 & 4.0&  51.7$\pm$ 3.2&  53.6$\pm$ 3.2&  1.47$\pm$0.04&  4.58$\pm$0.19&  0.86$\pm$0.09&  4.93$\pm$0.32&  39.8$\pm$ 2.7
\\ Togashi & 1.4& 11.601& 0.015& 2.0&  55.8$\pm$ 1.2&  30.3$\pm$ 1.4&  3.24$\pm$0.04&  4.70$\pm$0.07&  1.55$\pm$0.07&  5.53$\pm$0.35&  22.4$\pm$ 1.3
\\ Togashi & 1.4& 11.601& 0.015& 2.5&  52.6$\pm$ 1.7&  42.8$\pm$ 1.9&  2.30$\pm$0.04&  4.41$\pm$0.08&  1.03$\pm$0.06&  4.64$\pm$0.41&  32.1$\pm$ 2.1
\\ Togashi & 1.4& 11.601& 0.015& 3.0&  52.0$\pm$ 1.1&  48.4$\pm$ 1.3&  1.85$\pm$0.03&  4.32$\pm$0.08&  0.89$\pm$0.03&  4.23$\pm$0.34&  35.4$\pm$ 2.1
\\ Togashi & 1.4& 11.601& 0.015& 4.0&  51.6$\pm$ 2.0&  49.7$\pm$ 2.1&  1.38$\pm$0.03&  4.31$\pm$0.11&  0.87$\pm$0.05&  4.74$\pm$0.43&  37.2$\pm$ 2.1
\\ Togashi & 1.4& 11.601& 0.02 & 2.0&  57.4$\pm$ 1.0&  27.8$\pm$ 1.7&  3.06$\pm$0.06&  4.30$\pm$0.06&  1.55$\pm$0.10&  5.42$\pm$0.37&  20.1$\pm$ 2.0
\\ Togashi & 1.4& 11.601& 0.02 & 2.5&  53.0$\pm$ 0.9&  41.3$\pm$ 1.0&  2.19$\pm$0.01&  4.16$\pm$0.07&  1.01$\pm$0.02&  4.20$\pm$0.33&  31.3$\pm$ 1.7
\\ Togashi & 1.4& 11.601& 0.02 & 3.0&  52.6$\pm$ 3.7&  41.4$\pm$ 3.4&  1.82$\pm$0.08&  4.19$\pm$0.14&  1.02$\pm$0.12&  4.51$\pm$0.53&  30.5$\pm$ 2.2
\\ Togashi & 1.4& 11.601& 0.02 & 4.0&  52.1$\pm$ 1.4&  46.8$\pm$ 1.4&  1.30$\pm$0.02&  4.03$\pm$0.09&  0.86$\pm$0.03&  4.24$\pm$0.37&  34.2$\pm$ 2.1
\\ Togashi & 2.0& 11.231& 0.005& 2.0&  95.9$\pm$ 1.0&  40.2$\pm$ 0.5&  4.35$\pm$0.01&  5.77$\pm$0.06&  1.43$\pm$0.01&  5.04$\pm$0.27&  24.8$\pm$ 1.0
\\ Togashi & 2.0& 11.231& 0.005& 2.5&  95.7$\pm$ 1.0&  41.4$\pm$ 0.5&  3.55$\pm$0.02&  5.90$\pm$0.07&  1.43$\pm$0.01&  5.07$\pm$0.22&  25.5$\pm$ 1.1
\\ Togashi & 2.0& 11.231& 0.005& 3.0&  96.4$\pm$ 1.0&  43.4$\pm$ 0.5&  2.95$\pm$0.01&  5.83$\pm$0.07&  1.34$\pm$0.01&  5.07$\pm$0.28&  26.9$\pm$ 1.1
\\ Togashi & 2.0& 11.231& 0.005& 4.0&  98.7$\pm$ 2.6&  43.9$\pm$ 0.9&  2.14$\pm$0.03&  5.51$\pm$0.16&  1.26$\pm$0.05&  5.12$\pm$0.25&  28.0$\pm$ 1.3
\\ Togashi & 2.0& 11.231& 0.01 & 2.0&  97.8$\pm$ 1.8&  36.4$\pm$ 1.5&  3.58$\pm$0.02&  4.66$\pm$0.08&  1.28$\pm$0.06&  5.17$\pm$0.29&  24.7$\pm$ 1.3
\\ Togashi & 2.0& 11.231& 0.01 & 2.5&  96.7$\pm$ 2.2&  38.7$\pm$ 1.4&  2.95$\pm$0.02&  4.85$\pm$0.09&  1.25$\pm$0.05&  5.21$\pm$0.27&  25.3$\pm$ 1.2
\\ Togashi & 2.0& 11.231& 0.01 & 3.0&  96.6$\pm$ 1.3&  40.3$\pm$ 0.7&  2.50$\pm$0.01&  4.94$\pm$0.07&  1.22$\pm$0.01&  5.12$\pm$0.23&  26.1$\pm$ 1.2
\\ Togashi & 2.0& 11.231& 0.01 & 4.0&  97.9$\pm$ 1.2&  41.0$\pm$ 0.6&  1.89$\pm$0.01&  4.90$\pm$0.07&  1.19$\pm$0.01&  5.36$\pm$0.37&  27.0$\pm$ 1.1
\\ Togashi & 2.0& 11.231& 0.015& 2.0&  99.8$\pm$ 1.3&  34.6$\pm$ 0.5&  3.18$\pm$0.01&  4.05$\pm$0.06&  1.17$\pm$0.00&  5.20$\pm$0.28&  24.5$\pm$ 1.2
\\ Togashi & 2.0& 11.231& 0.015& 2.5&  97.8$\pm$ 1.6&  36.0$\pm$ 0.8&  2.61$\pm$0.01&  4.25$\pm$0.07&  1.18$\pm$0.02&  5.20$\pm$0.25&  24.6$\pm$ 1.3
\\ Togashi & 2.0& 11.231& 0.015& 3.0&  97.4$\pm$ 2.4&  38.0$\pm$ 1.2&  2.21$\pm$0.01&  4.34$\pm$0.09&  1.14$\pm$0.05&  5.23$\pm$0.30&  25.8$\pm$ 1.3
\\ Togashi & 2.0& 11.231& 0.015& 4.0&  97.9$\pm$ 2.5&  38.9$\pm$ 1.3&  1.68$\pm$0.01&  4.38$\pm$0.09&  1.13$\pm$0.05&  5.40$\pm$0.40&  25.8$\pm$ 1.1
\\ Togashi & 2.0& 11.231& 0.02 & 2.0& 101.2$\pm$ 1.7&  31.8$\pm$ 0.7&  2.91$\pm$0.01&  3.66$\pm$0.06&  1.15$\pm$0.02&  5.25$\pm$0.29&  23.3$\pm$ 1.1
\\ Togashi & 2.0& 11.231& 0.02 & 2.5&  99.5$\pm$ 1.6&  33.4$\pm$ 0.6&  2.39$\pm$0.01&  3.83$\pm$0.06&  1.15$\pm$0.02&  5.18$\pm$0.29&  23.7$\pm$ 1.2
\\ Togashi & 2.0& 11.231& 0.02 & 3.0&  98.3$\pm$ 2.3&  35.4$\pm$ 1.1&  2.03$\pm$0.01&  3.93$\pm$0.08&  1.11$\pm$0.05&  5.13$\pm$0.33&  24.7$\pm$ 1.3
\\ Togashi & 2.0& 11.231& 0.02 & 4.0&  98.3$\pm$ 2.2&  37.3$\pm$ 1.2&  1.54$\pm$0.01&  3.98$\pm$0.08&  1.07$\pm$0.05&  5.29$\pm$0.43&  25.2$\pm$ 1.4
\\ \hline TM1e & 1.4& 13.236& 0.005 & 2.0& --&--&--&--&--&--&--
\\ TM1e & 1.4& 13.236& 0.005& 2.5&  45.1$\pm$ 0.8&  41.7$\pm$ 2.0&  3.57$\pm$0.07&  6.90$\pm$0.13&  1.66$\pm$0.09&  5.69$\pm$0.37&  30.8$\pm$ 2.0
\\ TM1e & 1.4& 13.236& 0.005& 3.0&  42.9$\pm$ 0.9&  65.7$\pm$ 1.5&  2.44$\pm$0.03&  5.95$\pm$0.11&  0.91$\pm$0.03&  5.09$\pm$0.36&  47.4$\pm$ 2.1
\\ TM1e & 1.4& 13.236& 0.005& 4.0&  42.8$\pm$ 2.4&  67.4$\pm$ 2.9&  1.80$\pm$0.06&  5.88$\pm$0.23&  0.88$\pm$0.07&  5.12$\pm$0.60&  50.1$\pm$ 3.4
\\ TM1e & 1.4& 13.236& 0.01 & 2.0&  44.7$\pm$ 3.3&  50.5$\pm$ 8.8&  4.66$\pm$0.49&  7.27$\pm$0.61&  1.49$\pm$0.35&  5.57$\pm$0.73&  41.0$\pm$ 3.2
\\ TM1e & 1.4& 13.236& 0.01 & 2.5&  43.9$\pm$ 1.8&  51.5$\pm$ 2.7&  3.04$\pm$0.10&  6.03$\pm$0.19&  1.18$\pm$0.11&  4.54$\pm$0.41&  38.7$\pm$ 2.5
\\ TM1e & 1.4& 13.236& 0.01 & 3.0&  45.3$\pm$ 0.6&  38.8$\pm$ 0.4&  2.86$\pm$0.03&  6.59$\pm$0.07&  1.70$\pm$0.01&  6.06$\pm$0.32&  29.8$\pm$ 0.9
\\ TM1e & 1.4& 13.236& 0.01 & 4.0&  43.2$\pm$ 1.4&  60.5$\pm$ 1.8&  1.73$\pm$0.03&  5.59$\pm$0.15&  0.92$\pm$0.04&  5.06$\pm$0.39&  44.3$\pm$ 2.9
\\ TM1e & 1.4& 13.236& 0.015 & 2.0& --&--&--&--&--&--&--
\\ TM1e & 1.4& 13.236& 0.015 & 2.5& --&--&--&--&--&--&--
\\ TM1e & 1.4& 13.236& 0.015& 3.0&  46.2$\pm$ 0.7&  34.4$\pm$ 1.4&  2.70$\pm$0.05&  6.12$\pm$0.08&  1.78$\pm$0.06&  5.52$\pm$0.34&  28.9$\pm$ 1.9
\\ TM1e & 1.4& 13.236& 0.015& 4.0&  43.7$\pm$ 2.3&  54.2$\pm$ 2.4&  1.67$\pm$0.06&  5.31$\pm$0.18&  0.98$\pm$0.08&  5.02$\pm$0.58&  38.6$\pm$ 3.0
\\ TM1e & 1.4& 13.236& 0.02 & 2.0& --&--&--&--&--&--&--
\\ TM1e & 1.4& 13.236& 0.02 & 2.5& --&--&--&--&--&--&--
\\ TM1e & 1.4& 13.236& 0.02 & 3.0&  47.5$\pm$ 0.8&  30.8$\pm$ 0.8&  2.74$\pm$0.04&  6.04$\pm$0.09&  1.97$\pm$0.06&  5.47$\pm$0.09&  21.8$\pm$ 1.1
\\ TM1e & 1.4& 13.236& 0.02 & 4.0&  43.9$\pm$ 1.9&  47.6$\pm$ 2.1&  1.66$\pm$0.06&  5.29$\pm$0.18&  1.11$\pm$0.09&  4.98$\pm$0.47&  33.0$\pm$ 2.8
\\ TM1e & 2.0& 12.755& 0.005& 2.0&  78.5$\pm$ 1.9&  45.7$\pm$ 1.8&  4.66$\pm$0.05&  6.48$\pm$0.13&  1.42$\pm$0.07&  5.53$\pm$0.33&  32.7$\pm$ 1.8
\\ TM1e & 2.0& 12.755& 0.005& 2.5&  77.5$\pm$ 2.4&  50.7$\pm$ 2.4&  3.71$\pm$0.06&  6.52$\pm$0.15&  1.29$\pm$0.08&  5.39$\pm$0.33&  35.8$\pm$ 2.2
\\ TM1e & 2.0& 12.755& 0.005& 3.0&  76.7$\pm$ 1.7&  50.8$\pm$ 2.1&  3.12$\pm$0.03&  6.66$\pm$0.11&  1.31$\pm$0.06&  5.53$\pm$0.38&  35.3$\pm$ 1.9
\\ TM1e & 2.0& 12.755& 0.005& 4.0&  77.1$\pm$ 2.6&  53.9$\pm$ 2.0&  2.32$\pm$0.03&  6.58$\pm$0.16&  1.22$\pm$0.07&  5.54$\pm$0.32&  37.5$\pm$ 1.8
\\ TM1e & 2.0& 12.755& 0.01 & 2.0&  81.2$\pm$ 1.3&  41.6$\pm$ 1.2&  4.03$\pm$0.03&  5.41$\pm$0.09&  1.30$\pm$0.05&  5.42$\pm$0.37&  31.5$\pm$ 1.7
\\ TM1e & 2.0& 12.755& 0.01 & 2.5&  79.3$\pm$ 1.1&  42.5$\pm$ 1.1&  3.29$\pm$0.02&  5.65$\pm$0.07&  1.33$\pm$0.04&  5.30$\pm$0.31&  31.3$\pm$ 1.5
\\ TM1e & 2.0& 12.755& 0.01 & 3.0&  78.4$\pm$ 1.2&  47.0$\pm$ 2.0&  2.73$\pm$0.02&  5.69$\pm$0.07&  1.21$\pm$0.06&  5.01$\pm$0.44&  34.1$\pm$ 2.2
\\ TM1e & 2.0& 12.755& 0.01 & 4.0&  77.5$\pm$ 1.5&  48.9$\pm$ 1.2&  2.05$\pm$0.01&  5.78$\pm$0.10&  1.18$\pm$0.04&  5.21$\pm$0.37&  35.1$\pm$ 1.6
\\ TM1e & 2.0& 12.755& 0.015& 2.0&  83.6$\pm$ 1.5&  29.3$\pm$ 2.2&  3.93$\pm$0.07&  5.12$\pm$0.08&  1.76$\pm$0.12&  5.56$\pm$0.30&  21.2$\pm$ 2.0
\\ TM1e & 2.0& 12.755& 0.015& 2.5&  80.9$\pm$ 3.9&  38.0$\pm$ 3.5&  3.04$\pm$0.06&  5.14$\pm$0.16&  1.37$\pm$0.16&  5.24$\pm$0.35&  28.7$\pm$ 2.6
\\ TM1e & 2.0& 12.755& 0.015& 3.0&  78.9$\pm$ 1.4&  42.8$\pm$ 1.0&  2.52$\pm$0.01&  5.23$\pm$0.10&  1.22$\pm$0.03&  4.94$\pm$0.32&  31.5$\pm$ 1.4
\\ TM1e & 2.0& 12.755& 0.015& 4.0&  78.3$\pm$ 2.5&  44.3$\pm$ 1.3&  1.88$\pm$0.01&  5.24$\pm$0.14&  1.18$\pm$0.06&  5.12$\pm$0.36&  32.3$\pm$ 1.7
\\ TM1e & 2.0& 12.755& 0.02 & 2.0& --&--&--&--&--&--&--
\\ TM1e & 2.0& 12.755& 0.02 & 2.5&  82.6$\pm$ 3.3&  34.9$\pm$ 2.4&  2.85$\pm$0.04&  4.71$\pm$0.14&  1.36$\pm$0.13&  5.19$\pm$0.38&  27.4$\pm$ 2.6
\\ TM1e & 2.0& 12.755& 0.02 & 3.0&  79.6$\pm$ 1.4&  39.9$\pm$ 0.9&  2.36$\pm$0.01&  4.85$\pm$0.09&  1.22$\pm$0.02&  4.80$\pm$0.33&  29.6$\pm$ 1.5
\\ TM1e & 2.0& 12.755& 0.02 & 4.0&  78.2$\pm$ 2.2&  41.1$\pm$ 1.5&  1.76$\pm$0.01&  4.92$\pm$0.11&  1.20$\pm$0.05&  4.95$\pm$0.36&  30.0$\pm$ 1.5
\\
				\hline
			\end{tabular}
		}
		\end{center}
\end{table*}

\bibliography{ref}{}

\begin{thebibliography}{}
\expandafter\ifx\csname natexlab\endcsname\relax\def\natexlab#1{#1}\fi
\providecommand{\url}[1]{\href{#1}{#1}}
\providecommand{\dodoi}[1]{doi:~\href{http://doi.org/#1}{\nolinkurl{#1}}}
\providecommand{\doeprint}[1]{\href{http://ascl.net/#1}{\nolinkurl{http://ascl.net/#1}}}
\providecommand{\doarXiv}[1]{\href{https://arxiv.org/abs/#1}{\nolinkurl{https://arxiv.org/abs/#1}}}

\bibitem[{{Abbott} {et~al.}(2020{\natexlab{a}}){Abbott}, {Abbott}, {Abbott},
  {Abraham}, {Acernese}, \& {Virgo Collaboration}}]{2020ApJ...892L...3A}
{Abbott}, B.~P., {Abbott}, R., {Abbott}, T.~D., {et~al.} 2020{\natexlab{a}},
  \apjl, 892, L3, \dodoi{10.3847/2041-8213/ab75f5}

\bibitem[{{Abbott} {et~al.}(2020{\natexlab{b}}){Abbott}, {Abbott}, {Abbott},
  {Abraham}, {Acernese}, \& {Virgo Collaboration}}]{2020ApJ...896L..44A}
---. 2020{\natexlab{b}}, \apjl, 896, L44, \dodoi{10.3847/2041-8213/ab960f}

\bibitem[{{Abbott} {et~al.}(2018){Abbott}, {Abbott}, {Abbott}, {Acernese},
  {Ackley}, \& {Virgo Collaboration}}]{2018PhRvL.121p1101A}
---. 2018, \prl, 121, 161101, \dodoi{10.1103/PhysRevLett.121.161101}

\bibitem[{{Annala} {et~al.}(2018){Annala}, {Gorda}, {Kurkela}, \&
  {Vuorinen}}]{2018PhRvL.120q2703A}
{Annala}, E., {Gorda}, T., {Kurkela}, A., \& {Vuorinen}, A. 2018, \prl, 120,
  172703, \dodoi{10.1103/PhysRevLett.120.172703}

\bibitem[{{Arzoumanian} {et~al.}(2018){Arzoumanian}, {Brazier},
  {Burke-Spolaor}, {Chamberlin}, {Chatterjee}, {Christy}, {Cordes}, {Cornish},
  {Crawford}, {Thankful Cromartie}, {Crowter}, {DeCesar}, {Demorest}, {Dolch},
  {Ellis}, {Ferdman}, {Ferrara}, {Fonseca}, {Garver-Daniels}, {Gentile},
  {Halmrast}, {Huerta}, {Jenet}, {Jessup}, {Jones}, {Jones}, {Kaplan}, {Lam},
  {Lazio}, {Levin}, {Lommen}, {Lorimer}, {Luo}, {Lynch}, {Madison}, {Matthews},
  {McLaughlin}, {McWilliams}, {Mingarelli}, {Ng}, {Nice}, {Pennucci}, {Ransom},
  {Ray}, {Siemens}, {Simon}, {Spiewak}, {Stairs}, {Stinebring}, {Stovall},
  {Swiggum}, {Taylor}, {Vallisneri}, {van Haasteren}, {Vigeland}, {Zhu}, \&
  {NANOGrav Collaboration}}]{2018ApJS..235...37A}
{Arzoumanian}, Z., {Brazier}, A., {Burke-Spolaor}, S., {et~al.} 2018, \apjs,
  235, 37, \dodoi{10.3847/1538-4365/aab5b0}

\bibitem[{{Baiko} {et~al.}(2001){Baiko}, {Haensel}, \&
  {Yakovlev}}]{2001A&A...374..151B}
{Baiko}, D.~A., {Haensel}, P., \& {Yakovlev}, D.~G. 2001, \aap, 374, 151,
  \dodoi{10.1051/0004-6361:20010621}

\bibitem[{{Baym} {et~al.}(1971){Baym}, {Pethick}, \&
  {Sutherland}}]{1971ApJ...170..299B}
{Baym}, G., {Pethick}, C., \& {Sutherland}, P. 1971, \apj, 170, 299,
  \dodoi{10.1086/151216}

\bibitem[{{Belian} {et~al.}(1976){Belian}, {Conner}, \&
  {Evans}}]{1976ApJ...206L.135B}
{Belian}, R.~D., {Conner}, J.~P., \& {Evans}, W.~D. 1976, \apjl, 206, L135,
  \dodoi{10.1086/182151}

\bibitem[{{Blaschke} {et~al.}(2020){Blaschke}, {Ayriyan}, {Alvarez-Castillo},
  \& {Grigorian}}]{2020Univ....6...81B}
{Blaschke}, D., {Ayriyan}, A., {Alvarez-Castillo}, D.~E., \& {Grigorian}, H.
  2020, Universe, 6, 81, \dodoi{10.3390/universe6060081}

\bibitem[{{Brown}(2015)}]{2015ascl.soft05034B}
{Brown}, E.~F. 2015, {dStar: Neutron star thermal evolution code}.
\newblock \doeprint{1505.034}

\bibitem[{{Cromartie} {et~al.}(2020){Cromartie}, {Fonseca}, {Ransom},
  {Demorest}, {Arzoumanian}, {Blumer}, {Brook}, {DeCesar}, {Dolch}, {Ellis},
  {Ferdman}, {Ferrara}, {Garver-Daniels}, {Gentile}, {Jones}, {Lam}, {Lorimer},
  {Lynch}, {McLaughlin}, {Ng}, {Nice}, {Pennucci}, {Spiewak}, {Stairs},
  {Stovall}, {Swiggum}, \& {Zhu}}]{2020NatAs...4...72C}
{Cromartie}, H.~T., {Fonseca}, E., {Ransom}, S.~M., {et~al.} 2020, Nature
  Astronomy, 4, 72, \dodoi{10.1038/s41550-019-0880-2}

\bibitem[{{Cumming} {et~al.}(2006){Cumming}, {Macbeth}, {in 't Zand}, \&
  {Page}}]{2006ApJ...646..429C}
{Cumming}, A., {Macbeth}, J., {in 't Zand}, J.~J.~M., \& {Page}, D. 2006, \apj,
  646, 429, \dodoi{10.1086/504698}

\bibitem[{{Cyburt} {et~al.}(2010){Cyburt}, {Amthor}, {Ferguson}, {Meisel},
  {Smith}, {Warren}, {Heger}, {Hoffman}, {Rauscher}, {Sakharuk}, {Schatz},
  {Thielemann}, \& {Wiescher}}]{2010ApJS..189..240C}
{Cyburt}, R.~H., {Amthor}, A.~M., {Ferguson}, R., {et~al.} 2010, \apjs, 189,
  240, \dodoi{10.1088/0067-0049/189/1/240}

\bibitem[{{Deibel} {et~al.}(2016){Deibel}, {Meisel}, {Schatz}, {Brown}, \&
  {Cumming}}]{2016ApJ...831...13D}
{Deibel}, A., {Meisel}, Z., {Schatz}, H., {Brown}, E.~F., \& {Cumming}, A.
  2016, \apj, 831, 13, \dodoi{10.3847/0004-637X/831/1/13}

\bibitem[{{Demorest} {et~al.}(2010){Demorest}, {Pennucci}, {Ransom}, {Roberts},
  \& {Hessels}}]{2010Natur.467.1081D}
{Demorest}, P.~B., {Pennucci}, T., {Ransom}, S.~M., {Roberts}, M.~S.~E., \&
  {Hessels}, J.~W.~T. 2010, \nat, 467, 1081, \dodoi{10.1038/nature09466}

\bibitem[{{Dohi} {et~al.}(2020){Dohi}, {Hashimoto}, {Yamada}, {Matsuo}, \&
  {Fujimoto}}]{2020PTEP.2020c3E02D}
{Dohi}, A., {Hashimoto}, M.-a., {Yamada}, R., {Matsuo}, Y., \& {Fujimoto},
  M.~Y. 2020, Progress of Theoretical and Experimental Physics, 2020, 033E02,
  \dodoi{10.1093/ptep/ptaa010}

\bibitem[{{Dohi} {et~al.}(2019){Dohi}, {Nakazato}, {Hashimoto}, {Yasuhide}, \&
  {Noda}}]{2019PTEP.2019k3E01D}
{Dohi}, A., {Nakazato}, K., {Hashimoto}, M.-a., {Yasuhide}, M., \& {Noda}, T.
  2019, Progress of Theoretical and Experimental Physics, 2019, 113E01,
  \dodoi{10.1093/ptep/ptz116}

\bibitem[{{Fantina} {et~al.}(2018){Fantina}, {Zdunik}, {Chamel}, {Pearson},
  {Haensel}, \& {Goriely}}]{2018A&A...620A.105F}
{Fantina}, A.~F., {Zdunik}, J.~L., {Chamel}, N., {et~al.} 2018, \aap, 620,
  A105, \dodoi{10.1051/0004-6361/201833605}

\bibitem[{{Fujimoto} {et~al.}(1984){Fujimoto}, {Hanawa}, {Iben}, \&
  {Richardson}}]{1984ApJ...278..813F}
{Fujimoto}, M.~Y., {Hanawa}, T., {Iben}, I., J., \& {Richardson}, M.~B. 1984,
  \apj, 278, 813, \dodoi{10.1086/161851}

\bibitem[{{Fujimoto} {et~al.}(1981){Fujimoto}, {Hanawa}, \&
  {Miyaji}}]{1981ApJ...247..267F}
{Fujimoto}, M.~Y., {Hanawa}, T., \& {Miyaji}, S. 1981, \apj, 247, 267,
  \dodoi{10.1086/159034}

\bibitem[{{Fujimoto} {et~al.}(1987){Fujimoto}, {Sztajno}, {Lewin}, \& {van
  Paradijs}}]{1987ApJ...319..902F}
{Fujimoto}, M.~Y., {Sztajno}, M., {Lewin}, W. H.~G., \& {van Paradijs}, J.
  1987, \apj, 319, 902, \dodoi{10.1086/165507}

\bibitem[{{Fujimoto} {et~al.}(2003){Fujimoto}, {Hashimoto}, {Koike}, {Arai}, \&
  {Matsuba}}]{2003ApJ...585..418F}
{Fujimoto}, S.-i., {Hashimoto}, M.-a., {Koike}, O., {Arai}, K., \& {Matsuba},
  R. 2003, \apj, 585, 418, \dodoi{10.1086/345982}

\bibitem[{{Galloway} {et~al.}(2017){Galloway}, {Goodwin}, \&
  {Keek}}]{2017PASA...34...19G}
{Galloway}, D.~K., {Goodwin}, A.~J., \& {Keek}, L. 2017, \pasa, 34, e019,
  \dodoi{10.1017/pasa.2017.12}

\bibitem[{{Galloway} \& {Keek}(2021)}]{2021ASSL..461..209G}
{Galloway}, D.~K., \& {Keek}, L. 2021, Astrophysics and Space Science Library,
  461, 209, \dodoi{10.1007/978-3-662-62110-3_5}

\bibitem[{{Galloway} {et~al.}(2020){Galloway}, {in't Zand}, {Chenevez},
  {W{\"o}rpel}, {Keek}, {Ootes}, {Watts}, {Gisler}, {Sanchez-Fernandez}, \&
  {Kuulkers}}]{2020ApJS..249...32G}
{Galloway}, D.~K., {in't Zand}, J., {Chenevez}, J., {et~al.} 2020, \apjs, 249,
  32, \dodoi{10.3847/1538-4365/ab9f2e}

\bibitem[{{Grindlay} {et~al.}(1976){Grindlay}, {Gursky}, {Schnopper},
  {Parsignault}, {Heise}, {Brinkman}, \& {Schrijver}}]{1976ApJ...205L.127G}
{Grindlay}, J., {Gursky}, H., {Schnopper}, H., {et~al.} 1976, \apjl, 205, L127,
  \dodoi{10.1086/182105}

\bibitem[{{Haensel} \& {Zdunik}(1990)}]{1990A&A...227..431H}
{Haensel}, P., \& {Zdunik}, J.~L. 1990, \aap, 227, 431

\bibitem[{{Haensel} \& {Zdunik}(2008)}]{2008A&A...480..459H}
---. 2008, \aap, 480, 459, \dodoi{10.1051/0004-6361:20078578}

\bibitem[{{Heger} {et~al.}(2007{\natexlab{a}}){Heger}, {Cumming}, {Galloway},
  \& {Woosley}}]{2007ApJ...671L.141H}
{Heger}, A., {Cumming}, A., {Galloway}, D.~K., \& {Woosley}, S.~E.
  2007{\natexlab{a}}, \apjl, 671, L141, \dodoi{10.1086/525522}

\bibitem[{{Heger} {et~al.}(2007{\natexlab{b}}){Heger}, {Cumming}, \&
  {Woosley}}]{2007ApJ...665.1311H}
{Heger}, A., {Cumming}, A., \& {Woosley}, S.~E. 2007{\natexlab{b}}, \apj, 665,
  1311, \dodoi{10.1086/517491}

\bibitem[{{Johnston} {et~al.}(2020){Johnston}, {Heger}, \&
  {Galloway}}]{2020MNRAS.494.4576J}
{Johnston}, Z., {Heger}, A., \& {Galloway}, D.~K. 2020, \mnras, 494, 4576,
  \dodoi{10.1093/mnras/staa1054}

\bibitem[{{Keek} \& {Heger}(2011)}]{2011ApJ...743..189K}
{Keek}, L., \& {Heger}, A. 2011, \apj, 743, 189,
  \dodoi{10.1088/0004-637X/743/2/189}

\bibitem[{{Keek} {et~al.}(2012){Keek}, {Heger}, \& {in't
  Zand}}]{2012ApJ...752..150K}
{Keek}, L., {Heger}, A., \& {in't Zand}, J.~J.~M. 2012, \apj, 752, 150,
  \dodoi{10.1088/0004-637X/752/2/150}

\bibitem[{{Koike} {et~al.}(1999){Koike}, {Hashimoto}, {Arai}, \&
  {Wanajo}}]{1999A&A...342..464K}
{Koike}, O., {Hashimoto}, M., {Arai}, K., \& {Wanajo}, S. 1999, \aap, 342, 464

\bibitem[{{Koike} {et~al.}(2004){Koike}, {Hashimoto}, {Kuromizu}, \&
  {Fujimoto}}]{2004ApJ...603..242K}
{Koike}, O., {Hashimoto}, M.-a., {Kuromizu}, R., \& {Fujimoto}, S.-i. 2004,
  \apj, 603, 242, \dodoi{10.1086/381354}

\bibitem[{{Lalit} {et~al.}(2019){Lalit}, {Meisel}, \&
  {Brown}}]{2019ApJ...882...91L}
{Lalit}, S., {Meisel}, Z., \& {Brown}, E.~F. 2019, \apj, 882, 91,
  \dodoi{10.3847/1538-4357/ab338c}

\bibitem[{{Lampe} {et~al.}(2016){Lampe}, {Heger}, \&
  {Galloway}}]{2016ApJ...819...46L}
{Lampe}, N., {Heger}, A., \& {Galloway}, D.~K. 2016, \apj, 819, 46,
  \dodoi{10.3847/0004-637X/819/1/46}

\bibitem[{{Lattimer} \& {Swesty}(1991)}]{1991NuPhA.535..331L}
{Lattimer}, J.~M., \& {Swesty}, D.~F. 1991, \nphysa, 535, 331,
  \dodoi{10.1016/0375-9474(91)90452-C}

\bibitem[{{Lau} {et~al.}(2018){Lau}, {Beard}, {Gupta}, {Schatz}, {Afanasjev},
  {Brown}, {Deibel}, {Gasques}, {Hitt}, {Hix}, {Keek}, {M{\"o}ller},
  {Shternin}, {Steiner}, {Wiescher}, \& {Xu}}]{2018ApJ...859...62L}
{Lau}, R., {Beard}, M., {Gupta}, S.~S., {et~al.} 2018, \apj, 859, 62,
  \dodoi{10.3847/1538-4357/aabfe0}

\bibitem[{{Lim} {et~al.}(2017){Lim}, {Hyun}, \& {Lee}}]{2017IJMPE..2650015L}
{Lim}, Y., {Hyun}, C.~H., \& {Lee}, C.-H. 2017, International Journal of Modern
  Physics E, 26, 1750015, \dodoi{10.1142/S021830131750015X}

\bibitem[{{Liu} {et~al.}(2017){Liu}, {Matsuo}, {Hashimoto}, {Noda}, \&
  {Fujimoto}}]{2017JPSJ...86l3901L}
{Liu}, H., {Matsuo}, Y., {Hashimoto}, M.-a., {Noda}, T., \& {Fujimoto}, M.~Y.
  2017, Journal of the Physical Society of Japan, 86, 123901,
  \dodoi{10.7566/JPSJ.86.123901}

\bibitem[{{Matsuo}(2017)}]{2017KUPhD1806813}
{Matsuo}, Y. 2017, PhD thesis in Kyushu Univ., \dodoi{10.15017/1806813}

\bibitem[{{Meisel}(2018)}]{2018ApJ...860..147M}
{Meisel}, Z. 2018, \apj, 860, 147, \dodoi{10.3847/1538-4357/aac3d3}

\bibitem[{{Meisel} {et~al.}(2019){Meisel}, {Merz}, \&
  {Medvid}}]{2019ApJ...872...84M}
{Meisel}, Z., {Merz}, G., \& {Medvid}, S. 2019, \apj, 872, 84,
  \dodoi{10.3847/1538-4357/aafede}

\bibitem[{{Miller} {et~al.}(2019){Miller}, {Lamb}, {Dittmann}, {Bogdanov},
  {Arzoumanian}, {Gendreau}, {Guillot}, {Harding}, {Ho}, {Lattimer}, {Ludlam},
  {Mahmoodifar}, {Morsink}, {Ray}, {Strohmayer}, {Wood}, {Enoto}, {Foster},
  {Okajima}, {Prigozhin}, \& {Soong}}]{2019ApJ...887L..24M}
{Miller}, M.~C., {Lamb}, F.~K., {Dittmann}, A.~J., {et~al.} 2019, \apjl, 887,
  L24, \dodoi{10.3847/2041-8213/ab50c5}

\bibitem[{{Most} {et~al.}(2018){Most}, {Weih}, {Rezzolla}, \&
  {Schaffner-Bielich}}]{2018PhRvL.120z1103M}
{Most}, E.~R., {Weih}, L.~R., {Rezzolla}, L., \& {Schaffner-Bielich}, J. 2018,
  \prl, 120, 261103, \dodoi{10.1103/PhysRevLett.120.261103}

\bibitem[{{Noda} {et~al.}(2013){Noda}, {Hashimoto}, {Yasutake}, {Maruyama},
  {Tatsumi}, \& {Fujimoto}}]{2013ApJ...765....1N}
{Noda}, T., {Hashimoto}, M.-a., {Yasutake}, N., {et~al.} 2013, \apj, 765, 1,
  \dodoi{10.1088/0004-637X/765/1/1}

\bibitem[{{Page} {et~al.}(2013){Page}, {Lattimer}, {Prakash}, \&
  {Steiner}}]{2013arXiv1302.6626P}
{Page}, D., {Lattimer}, J.~M., {Prakash}, M., \& {Steiner}, A.~W. 2013, arXiv
  e-prints, arXiv:1302.6626.
\newblock \doarXiv{1302.6626}

\bibitem[{{Page} {et~al.}(2011){Page}, {Prakash}, {Lattimer}, \&
  {Steiner}}]{2011PhRvL.106h1101P}
{Page}, D., {Prakash}, M., {Lattimer}, J.~M., \& {Steiner}, A.~W. 2011, \prl,
  106, 081101, \dodoi{10.1103/PhysRevLett.106.081101}

\bibitem[{{Parikh} {et~al.}(2013){Parikh}, {Jos{\'e}}, {Sala}, \&
  {Iliadis}}]{2013PrPNP..69..225P}
{Parikh}, A., {Jos{\'e}}, J., {Sala}, G., \& {Iliadis}, C. 2013, Progress in
  Particle and Nuclear Physics, 69, 225, \dodoi{10.1016/j.ppnp.2012.11.002}

\bibitem[{{Potekhin} {et~al.}(2015){Potekhin}, {Pons}, \&
  {Page}}]{2015SSRv..191..239P}
{Potekhin}, A.~Y., {Pons}, J.~A., \& {Page}, D. 2015, \ssr, 191, 239,
  \dodoi{10.1007/s11214-015-0180-9}

\bibitem[{{Raaijmakers} {et~al.}(2019){Raaijmakers}, {Riley}, {Watts}, {Greif},
  {Morsink}, {Hebeler}, {Schwenk}, {Hinderer}, {Nissanke}, {Guillot},
  {Arzoumanian}, {Bogdanov}, {Chakrabarty}, {Gendreau}, {Ho}, {Lattimer},
  {Ludlam}, \& {Wolff}}]{2019ApJ...887L..22R}
{Raaijmakers}, G., {Riley}, T.~E., {Watts}, A.~L., {et~al.} 2019, \apjl, 887,
  L22, \dodoi{10.3847/2041-8213/ab451a}

\bibitem[{{Richardson} {et~al.}(1982){Richardson}, {van Horn}, {Ratcliff}, \&
  {Malone}}]{1982ApJ...255..624R}
{Richardson}, M.~B., {van Horn}, H.~M., {Ratcliff}, K.~F., \& {Malone}, R.~C.
  1982, \apj, 255, 624, \dodoi{10.1086/159865}

\bibitem[{{Schatz} {et~al.}(1999){Schatz}, {Bildsten}, {Cumming}, \&
  {Wiescher}}]{1999ApJ...524.1014S}
{Schatz}, H., {Bildsten}, L., {Cumming}, A., \& {Wiescher}, M. 1999, \apj, 524,
  1014, \dodoi{10.1086/307837}

\bibitem[{{Schatz} \& {Rehm}(2006)}]{2006NuPhA.777..601S}
{Schatz}, H., \& {Rehm}, K.~E. 2006, \nphysa, 777, 601,
  \dodoi{10.1016/j.nuclphysa.2005.05.200}

\bibitem[{{Schatz} {et~al.}(1998){Schatz}, {Aprahamian}, {Goerres}, {Wiescher},
  {Rauscher}, {Rembges}, {Thielemann}, {Pfeiffer}, {Moeller}, {Kratz},
  {Herndl}, {Brown}, \& {Rebel}}]{1998PhR...294..167S}
{Schatz}, H., {Aprahamian}, A., {Goerres}, J., {et~al.} 1998, \physrep, 294,
  167, \dodoi{10.1016/S0370-1573(97)00048-3}

\bibitem[{{Schatz} {et~al.}(2001){Schatz}, {Aprahamian}, {Barnard}, {Bildsten},
  {Cumming}, {Ouellette}, {Rauscher}, {Thielemann}, \&
  {Wiescher}}]{2001PhRvL..86.3471S}
{Schatz}, H., {Aprahamian}, A., {Barnard}, V., {et~al.} 2001, \prl, 86, 3471,
  \dodoi{10.1103/PhysRevLett.86.3471}

\bibitem[{{Shchechilin} {et~al.}(2021){Shchechilin}, {Gusakov}, \&
  {Chugunov}}]{2021arXiv210501991S}
{Shchechilin}, N.~N., {Gusakov}, M.~E., \& {Chugunov}, A.~I. 2021, arXiv
  e-prints, arXiv:2105.01991.
\newblock \doarXiv{2105.01991}

\bibitem[{{Shen} {et~al.}(2020){Shen}, {Ji}, {Hu}, \&
  {Sumiyoshi}}]{2020ApJ...891..148S}
{Shen}, H., {Ji}, F., {Hu}, J., \& {Sumiyoshi}, K. 2020, \apj, 891, 148,
  \dodoi{10.3847/1538-4357/ab72fd}

\bibitem[{{Shen} {et~al.}(1998{\natexlab{a}}){Shen}, {Toki}, {Oyamatsu}, \&
  {Sumiyoshi}}]{1998NuPhA.637..435S}
{Shen}, H., {Toki}, H., {Oyamatsu}, K., \& {Sumiyoshi}, K. 1998{\natexlab{a}},
  \nphysa, 637, 435, \dodoi{10.1016/S0375-9474(98)00236-X}

\bibitem[{{Shen} {et~al.}(1998{\natexlab{b}}){Shen}, {Toki}, {Oyamatsu}, \&
  {Sumiyoshi}}]{1998PThPh.100.1013S}
---. 1998{\natexlab{b}}, Progress of Theoretical Physics, 100, 1013,
  \dodoi{10.1143/PTP.100.1013}

\bibitem[{{Shen} {et~al.}(2011){Shen}, {Toki}, {Oyamatsu}, \&
  {Sumiyoshi}}]{2011ApJS..197...20S}
---. 2011, \apjs, 197, 20, \dodoi{10.1088/0067-0049/197/2/20}

\bibitem[{{Shternin} {et~al.}(2011){Shternin}, {Yakovlev}, {Heinke}, {Ho}, \&
  {Patnaude}}]{2011MNRAS.412L.108S}
{Shternin}, P.~S., {Yakovlev}, D.~G., {Heinke}, C.~O., {Ho}, W. C.~G., \&
  {Patnaude}, D.~J. 2011, \mnras, 412, L108,
  \dodoi{10.1111/j.1745-3933.2011.01015.x}

\bibitem[{{Sugimoto} \& {Fujimoto}(1978)}]{1978PASJ...30..467S}
{Sugimoto}, D., \& {Fujimoto}, M.~Y. 1978, \pasj, 30, 467

\bibitem[{{Sugimoto} {et~al.}(1981){Sugimoto}, {Nomoto}, \&
  {Eriguchi}}]{1981PThPS..70..115S}
{Sugimoto}, D., {Nomoto}, K., \& {Eriguchi}, Y. 1981, Progress of Theoretical
  Physics Supplement, 70, 115, \dodoi{10.1143/PTPS.70.115}

\bibitem[{{Sumiyoshi} {et~al.}(2019){Sumiyoshi}, {Nakazato}, {Suzuki}, {Hu}, \&
  {Shen}}]{2019ApJ...887..110S}
{Sumiyoshi}, K., {Nakazato}, K., {Suzuki}, H., {Hu}, J., \& {Shen}, H. 2019,
  \apj, 887, 110, \dodoi{10.3847/1538-4357/ab5443}

\bibitem[{{Taam}(1980)}]{1980ApJ...241..358T}
{Taam}, R.~E. 1980, \apj, 241, 358, \dodoi{10.1086/158348}

\bibitem[{{Tanaka}(1989)}]{1989ESASP.296....3T}
{Tanaka}, Y. 1989, in ESA Special Publication, Vol.~1, Two Topics in X-Ray
  Astronomy, Volume 1: X Ray Binaries. Volume 2: AGN and the X Ray Background,
  ed. J.~{Hunt} \& B.~{Battrick}, 3

\bibitem[{{Togashi} {et~al.}(2017){Togashi}, {Nakazato}, {Takehara},
  {Yamamuro}, {Suzuki}, \& {Takano}}]{2017NuPhA.961...78T}
{Togashi}, H., {Nakazato}, K., {Takehara}, Y., {et~al.} 2017, \nphysa, 961, 78,
  \dodoi{10.1016/j.nuclphysa.2017.02.010}

\bibitem[{{Ubertini} {et~al.}(1999){Ubertini}, {Bazzano}, {Cocchi},
  {Natalucci}, {Heise}, {Muller}, \& {in 't Zand}}]{1999ApJ...514L..27U}
{Ubertini}, P., {Bazzano}, A., {Cocchi}, M., {et~al.} 1999, \apjl, 514, L27,
  \dodoi{10.1086/311933}

\bibitem[{{Woosley} {et~al.}(2004){Woosley}, {Heger}, {Cumming}, {Hoffman},
  {Pruet}, {Rauscher}, {Fisker}, {Schatz}, {Brown}, \&
  {Wiescher}}]{2004ApJS..151...75W}
{Woosley}, S.~E., {Heger}, A., {Cumming}, A., {et~al.} 2004, \apjs, 151, 75,
  \dodoi{10.1086/381533}

\bibitem[{{Yakovlev} {et~al.}(2001){Yakovlev}, {Kaminker}, {Gnedin}, \&
  {Haensel}}]{2001PhR...354....1Y}
{Yakovlev}, D.~G., {Kaminker}, A.~D., {Gnedin}, O.~Y., \& {Haensel}, P. 2001,
  \physrep, 354, 1, \dodoi{10.1016/S0370-1573(00)00131-9}

\bibitem[{{Yakovlev} \& {Pethick}(2004)}]{2004ARA&A..42..169Y}
{Yakovlev}, D.~G., \& {Pethick}, C.~J. 2004, \araa, 42, 169,
  \dodoi{10.1146/annurev.astro.42.053102.134013}

\end{thebibliography}
\bibliographystyle{aasjournal}

\end{document}